\documentclass[aps,prb,twocolumn,superscriptaddress]{revtex4}
\usepackage{graphics}
\usepackage{graphicx}
\usepackage[english]{babel}
\usepackage{color}
\usepackage{amsmath}
\usepackage{mathtools}

\newcommand{\Bi}[1]{Bi$_{2}$Sr$_{2}$CaCu$_{2}$O$_{8 + \delta}$\,\,}

\begin{document}

\title{Unveiling the vortex glass phase in the surface and volume of
a type-II superconductor }

\author{Jazm\'{i}n Arag\'{o}n S\'{a}nchez}
\affiliation{Centro At\'omico Bariloche and Instituto Balseiro,
CNEA, 8400 Bariloche, Argentina.}
\author{Ra\'ul Cort\'es Maldonado}
\affiliation{Centro At\'omico Bariloche and Instituto Balseiro,
CNEA, 8400 Bariloche, Argentina.}
\author{N\'estor Ren\'e Cejas Bolecek}
\affiliation{Centro At\'omico Bariloche and Instituto Balseiro,
CNEA, 8400 Bariloche, Argentina.}
\author{Gonzalo Rumi}
\affiliation{Centro At\'omico Bariloche and Instituto Balseiro,
CNEA, 8400 Bariloche, Argentina.}
\author{Pablo Pedrazzini}
\affiliation{Centro At\'omico Bariloche and Instituto Balseiro,
CNEA, 8400 Bariloche, Argentina.}
\author{Moira I. Dolz}
\affiliation{Universidad Nacional de San Luis and Instituto de
F\'{i}sica Aplicada, CONICET, 5700 San Luis, Argentina.}
\author{Gladys Nieva}
\affiliation{Centro At\'omico Bariloche and Instituto Balseiro,
CNEA, 8400 Bariloche, Argentina.}
\author{Cornelis J. van der Beek}
\affiliation{Laboratoire des Solides Irradi\'{e}s, \'{E}cole
Polytechnique, CNRS, 91128 Palaiseau, France.}
\author{Marcin Konczykowski}
\affiliation{Laboratoire des Solides Irradi\'{e}s, \'{E}cole
Polytechnique, CNRS, 91128 Palaiseau, France.}
\author{C. D. Dewhurst}
\affiliation{Institut Laue-Langevin, B.P. 156, 38042 Grenoble Cedex
9, France}
\author{R. Cubitt}
\affiliation{Institut Laue-Langevin, B.P. 156, 38042 Grenoble Cedex
9, France}
\author{Alejandro B. Kolton}
\affiliation{Centro At\'omico Bariloche and Instituto Balseiro,
CNEA, 8400 Bariloche, Argentina.}
\author{Alain Pautrat}
\affiliation{Laboratoire CRISMAT-EnsiCaen, Caen, France.}
\author{Yanina Fasano$^{*}$}
\affiliation{Centro At\'omico Bariloche and Instituto Balseiro,
CNEA, 8400 Bariloche, Argentina.}

\date{\today}

\begin{abstract}

Order-disorder transitions between glassy phases are quite common in
nature and yet a comprehensive survey of the microscopic structural
changes remains elusive since the scale of the constituents is tiny
and in most cases few of them take part in the transformation.
Vortex matter in type-II superconductors is a model system where
some of the experimental challenges inherent to this general
question can be tackled by adequately choosing the host
superconducting sample. For instance,
Bi$_{2}$Sr$_{2}$CaCu$_{2}$O$_{8 + \delta}$ is a type-II
superconductor with weak point disorder that presents a transition
between two glassy phases on increasing the constituents' (vortices)
density. At low vortex densities, the impact of disorder produces
the nucleation of a glassy yet quasi-crystalline phase, the Bragg
glass. For high vortex densities the stable phase, coined as
\textit{vortex glass}, was proposed to be disordered, but its
structural properties have remained elusive up to now. Here we
answer this question by combining surface and bulk vortex imaging
techniques, and show that the vortex glass is neither a messy nor a
hexatic phase: in the plane of vortices it presents large
crystallites with positional correlations growing algebraically and
short-ranged orientational order. However, no dramatic change in the
correlation length along the direction of vortices is observed on
traversing the order-disorder transformation.

\end{abstract}

\keywords{order-disorder transitions $|$ glass-to-glass
transformation $|$ single-particle imaging in structural
transformations$|$ vortex glass phase $|$ type-II superconductors }

\maketitle

Order-disorder structural transitions are such ubiquitous in nature
that we are faced to solid-to-liquid and solid-to-solid structural
transformations in everyday experience. Quite frequently, the
unavoidable disorder present in the materials that are transforming
favors the stabilization of glassy phases, such as spin, electric,
and superconducting glass systems.~\cite{Mydosh} In order to expand
the potential applications of these materials, we need to have
direct access, with microscopic resolution, on the structural
changes entailed at order-disorder transformations. Considerable
experimental information on this issue was obtained in the
solid-to-liquid case, concerning monolayers of electrons trapped on
the surface of liquid He,~\cite{Grimes1979} colloidal and hard
spheres,~\cite{Murray1987,Marcus1996,Thorneywork2017} plasma
crystals,~\cite{Thomas1996} and superparamagnetic colloidal
particles~\cite{Deutschlander2013}. These experiments support the
dislocation-mediated two-stage melting with an intermediate hexatic
phase proposed by renowned theoretical developments in the case of
melting in two dimensions,
~\cite{Kosterlitz1973,Halperin1978,Nelson1979,Young1979,Chui1983}
spin,~\cite{Kosterlitz1973} and superfluid
transitions.~\cite{Kosterlitz1974} Only a handful of studies have
directly probed the structural changes entailed in glass-to-glass
order-disorder transitions in three dimensions.

Vortex matter in type-II superconductors is a model condensed matter
system to study this general problem since the energy scales and
dimensionality  can be easily tuned by control parameters. The three
competing energy scales governing the occurrence of order-disorder
transformations are vortex-vortex repulsion, vortex-pinning
interaction and thermal fluctuations, respectively controlled by
magnetic field $H$, sample disorder, and temperature. Order-disorder
transitions in vortex matter occur since the vortex-vortex
interaction tends to form an ordered triangular lattice whereas
thermal fluctuations and the pinning produced by disorder in the
crystal structure of the sample conspire against the stabilization
of a perfect crystal.~\cite{Blatter} Some experimental works apply
imaging techniques to reveal the  structural changes entailed in
glass-to-glass transitions in vortex
matter,~\cite{Pardo1998,Troyanovsky1999,Menghini2002,Fasano2002,Petrovic2009,Guillamon2014,Zehetmayer2015,Ganguli2015,Toft-Petersen2018}
but focus on the in-plane microscopic details, i.e.,  at the sample
surface. Other works study the structural changes along the third
direction only and found that at the vortex
melting~\cite{Shibauchi1999} and glass-to-glass~\cite{Gaifullin2000}
transitions the coupling  of vortices along the c-axis is depleted
in the disordered phases.  In order to have a comprehensive
description of the problem, studies on imaging the in-plane as well
as out-of-plane structural changes taking place in order-disorder
transitions in vortex matter are required. Experimentally, this is
quite challenging since requires the ability to image, with
single-particle resolution, vortex matter at the surface and bulk of
the sample. This is the question we are dealing with here.

In some superconducting glasses, as in the case of vortex matter
nucleated in the extremely-layered high-$T_{\rm c}$
Bi$_{2}$Sr$_{2}$CaCu$_{2}$O$_{8 + \delta}$ that we study here,
 a quasi-crystalline state of matter is stable at low vortex
 densities. This phase known as the Bragg
glass presents a weak logarithmic decay of positional order and
glassy dynamics.~\cite{Giamarchi1995,Klein2001} On increasing vortex
density (magnetic field), a first-order transition towards another
phase, the \textit{vortex glass}, takes place at a characteristic
field $B_{\rm ord-dis}$.~\cite{Natterman2000} The vortex glass is
expected to present a positional order that is not
quasi-crystalline, with non-divergent peaks in the structure factor
as in the case of the Bragg glass. The order-disorder transition
field is frequently observed as concomitant to the onset of the
so-called \textit{second-peak
effect}.~\cite{Khaykovich97a,Vinokur1998,Cubbit1993,Avraham2001}
This effect refers to the sudden enhancement or peak in the
superconductor critical current when increasing field or
temperature, as a consequence of pinning energy overcoming the
elastic energy in the disordered phase. The $B_{\rm ord-dis}$
transition field at which the vortex density presents a
jump~\cite{Avraham2001} coincides with the onset field of the second
peak effect, $B_{\rm on}$. At high temperatures the order-disorder
transition continues as a first-order melting at the $B_{\rm FOT}$
line separating the Bragg glass and the vortex liquid
phase.~\cite{Pastoriza1994} The phase coherence along the direction
of vortices is depleted both at $B_{\rm ord-dis}$ and $B_{\rm
FOT}$,~\cite{Gaifullin2000} namely at the whole border of the Bragg
glass phase.

Small angle neutron scattering (SANS) is a bulk-sensitive technique
that allows to extract quantitative information on real space
correlation lengths. For the model superconducting glass we study
here, previous SANS data reveal that the vortex diffraction pattern
does not significantly change on traversing the $B_{\rm ord-dis}$,
~\cite{Forgan1996,Pautrat2007}, but the scattered intensity was
strongly reduced for $B > B_{\rm ord-dis}$. Therefore, precise
details on the structural properties of the vortex glass phase were
not accessible in those works. A similar resolution-limited problem
was also reported for SANS data in the vortex glass phase of other
superconductors.~\cite{Klein2001,Demirdis2016} Therefore, SANS
experiments with improved resolution are needed to characterize the
fine structural changes entailed at the glass-to-glass
transformation. In this work we combine such SANS measurements with
direct imaging of the Bi$_{2}$Sr$_{2}$CaCu$_{2}$O$_{8 + \delta}$
vortex structure in large fields-of-view at the surface of the
sample for the ordered and disordered phases. We unveil the
evolution of the in-plane and longitudinal vortex correlation
lengths, as well as distance-evolution of the in-plane orientational
and positional order, on traversing the Bragg-to-vortex glass
transition.

\section*{Results}

\subsection{Effect of point disorder on the Bragg-to-vortex glass transition}

The order-disorder transition in pristine
Bi$_{2}$Sr$_{2}$CaCu$_{2}$O$_{8 + \delta}$ vortex matter occurs at
approximately 200-300\,Gauss, well above the vortex density range
that can be probed by most real-space techniques that grant access
to wide fields-of-view of thousands of vortices.~\cite{Fasano2008}
However, introducing extra point disorder in the samples by
irradiation with electrons at low temperatures  results in a
significant lowering of the order-disorder line in the phase
diagram.~\cite{Khaykovich97a,Konczykowski2009} Such a procedure
allows us to apply the magnetic decoration technique
(MD)~\cite{Fasano2005} to observe, in real space, and over large
fields-of-view, the structural changes entailed at the surface of
these samples at $B_{\rm ord-dis}$.

Our samples are pristine and electron-irradiated
Bi$_{2}$Sr$_{2}$CaCu$_{2}$O$_{8 + \delta}$ single crystals. During
irradiation, high-energy electrons traverse the entire sample
thickness, generating an almost homogeneous distribution of atomic
point defects: roughly 10\,\% of atoms, depending on the irradiation
dose, may be displaced from its initial position. The pristine
samples P studied by MD are 30 small single crystals, whereas for
SANS measurements we used one millimetric-large single crystal. We
also study two electron-irradiated samples, A and B, first
oxygen-annealed and then irradiated with 2.3\,MeV electrons at
20\,K, which produced a significant decrease of  $B_{\rm
ord-dis}$.~\cite{Khaykovich97a} Sample B has a larger irradiation
dose than sample A, see Methods section. The changes in the vortex
phase diagram induced by irradiation with electrons, are tracked by
means of dc and ac local Hall magnetometry. Detailed data on the
detection of $B_{\rm FOT}$, $B_{\rm ord-dis}=B_{\rm ON}$, and the
sweep-rate dependent $B_{\rm SP}$ field at which the critical
current presents a peak on increasing $H$ can be found in the
Supplementary Material section.

Figure \ref{Figure1} shows the vortex phase diagram for pristine and
electron-irradiated Bi$_{2}$Sr$_{2}$CaCu$_{2}$O$_{8 + \delta}$
samples A and B. Increasing the dose of irradiation with electrons
produces a systematic decrease of $T_{\rm c}$ and $B_{\rm
ord-dis}$,~\cite{Konczykowski2009} and then the Bragg glass phase
spans a smaller $B\,-\,T$ phase region. At low temperatures, sample
A has a $B_{\rm ord-dis} = 85 (-5)$\,Gauss and for sample B this
value is even reduced to $\sim 40 (-8)$\,Gauss. The order-disorder
transition field also depends on the oxygen-doping and annealing of
the samples,~\cite{Correa2001,Khaykovich97a} and this is the reason
why sample B presents the  smallest reported value of $B_{\rm
ord-dis}$ obtained by electron irradiation. This  allowed us to
reveal the structural properties of the vortex glass phase in
extended fields-of-view by means of MD.

\subsection{Bragg-to-vortex glass transformation from real-space MD imaging}

Figure\,\ref{Figure2} shows MD snapshots with more than 1500
individually-resolved vortices taken after field-cooling the samples
down to 4.2\,K, at different applied fields above and below $B_{\rm
SP}$. Changing field varies the lattice spacing of the hexagonal
vortex lattice $a_{0}=1.075 \sqrt{\Phi_{0}/B}$. Vortices are
decorated with Fe particles attracted by the local field gradient
generated around the cores, observed as the black dots in the
inverted scanning-electron-microscopy images of Fig.\,\ref{Figure2}.
The decorated structures were frozen, at length scales of $a_{0}$,
at temperatures at which the pinning generated by disorder sets
in.~\cite{Pardo1997,Fasano2005} This freezing temperature is some K
below the temperature at which magnetic response becomes
irreversible,~\cite{CejasBolecek2016} see the $B_{\rm IL}$ in
Fig.\,\ref{Figure1} (dashed lines). For $B \le B_{\rm ord-dis}$ the
$B_{\rm IL}$ line coincides with the melting line $B_{\rm FOT}$
within the experimental error.

Panels (a) and (b) of Fig.\,\ref{Figure2} correspond to snapshots of
vortex positions taken at $B<B_{\rm ord-dis}$ for the Bragg glass
phase. For fields larger than 15\,Gauss, the vortex structure is
single-crystalline and presents very few topological defects
associated with non-sixfold coordinated vortices. For instance, the
vortex structure of Fig.\,\ref{Figure2} (a) nucleated in a pristine
sample for  $B/B_{\rm ord-dis} = 0.28$ is defectless in this
field-of-view. For fields smaller than 15\,Gauss, the structure
breaks into small crystallites for pristine~\cite{Fasano1999} as
well as electron-irradiated samples. This polycrystalline structure
results from vortex-vortex interaction weakening and disorder
becoming more relevant on the viscous freezing
dynamics.~\cite{Fasano1999}

Vortices are better resolved by MD in pristine than in
electron-irradiated samples: at a given $B$, the
maximum-to-background intensity of the decoration image is larger in P samples.
 This is in agreement with the
reported increase in the superconducting penetration depth
$\lambda(0)$ when irradiating with
electrons.~\cite{Konczykowski2009} For samples A and B,  we estimate
$\lambda(0)$ is 30\% larger than in P samples from the difference in
the entropy-jump at $B_{\rm FOT}$.~\cite{Konczykowski2009}  An
enhancement of $\lambda(0)$ results in a decrease of the local field
gradient that diminishes the magnetic force attracting the Fe
particles that decorate individual vortex positions. This reduces
the maximum field up to which individual vortices can be resolved
with MD in electron-irradiated samples, $\sim 90$\,Gauss for samples
A and B.

Nevertheless, for sample B this field is high enough as to take
snapshots of the vortex glass phase in extended fields-of-view.
Figure \ref{Figure2} shows the largest picture of the vortex glass
phase that we have obtained for $B/B_{\rm ord-dis}=1.62$,  with more
than 1500 vortices. The overimposed Delaunay triangulation joining
first-neighbor vortices highlights the presence of grain boundaries
separating relatively large crystallites with hundreds of vortices.
The orientation between crystallites changes between 20 to 30
degrees. Roughly 6\% of vortices are non-sixfold coordinated and
form different topological defects: 60\% participate in grain
boundaries (violet triangles), 19\% form isolated edge dislocations
(orange triangles) and the remaining 21\% are involved in
twisted-bond deformations (pink triangles). Isolated edge
dislocations are plastic deformations entailing the nucleation of
two extra vortex planes; twisted bonds are two adjacent edge
dislocations with opposite Burgers vectors and do not entail any
extra vortex plane. Twisted bonds are local elastic deformations
that can be cured by displacing individual vortex positions in a
fraction of the lattice spacing. This snapshot of the vortex glass
reveals that, within a crystallite, isolated edge dislocations
entail plastic deformations even at large length scales. Indeed,
only $\sim 40$\,\% of the observed edge dislocations are paired with
their Burgers vectors pointing in opposite directions (red arrows in
Fig.\,\ref{Figure2} (c)).

 The proliferation of topological defects in the vortex glass phase
 contrasts with most of the Bragg glass phase being single-crystalline with
 a fraction of non-sixfold coordinated vortices $\rho_{\rm def}<1$\,\%.
 Figure \,\ref{Figure3} (a) shows
the evolution of this fraction, $\rho_{\rm def}$, as a function of $
B/B_{\rm ord-dis}$ for samples A, B and P. At low fields $B/B_{\rm
ord-dis} \lesssim 0.01$, the vortex structure fractures into small
crystallites and $\rho_{\rm def} \sim 50$\,\%. On increasing field,
$\rho_{\rm def}$ decays dramatically and stagnates below 1\,\% for
$B/B_{\rm ord-dis} \gtrsim 0.1$, in concomitance with the
observation of a single-crystalline vortex structure. This
phenomenology is common to P and electron-irradiated samples, but
the precise $B/B_{\rm ord-dis}$ value at which the structure becomes
single-crystalline depends on the magnitude of point disorder. For
instance, for $B/B_{\rm ord-dis}=0.09$ the vortex structure
nucleated in sample A has large crystallites whereas that in sample
P is single-crystalline, see the structure factor patterns of
Fig.\,\ref{Figure3} (b). These patterns result from calculating the
structure factor from the individual vortex positions detected at
the sample surface, namely $S_{\rm MD}(q,\Psi)= \mid
\tilde{\rho}(q_{\rm x},q_{\rm y})\mid ^{2}_{z=0}$, with
$\tilde{\rho}$ the Fourier transform  of the density of vortex lines
$\rho(x,y,z)=\frac{1}{t} \Sigma_{j=1}^{N} \delta(x - x_{\rm
j}(z))\delta(y - y_{\rm j}(z))$ with $t$ the thickness of the sample
and $N$ the number of vortices. On increasing field above $B_{\rm
ord-dis}$, well within the vortex glass phase, the density of
defects enhances up to 6.5\% at $B/B_{\rm ord-dis}=1.62$. Within our
experimental field-of-view the structure presents four large
crystallites, resulting in multiple Bragg peaks in the $S_{\rm
MD}(q,\Psi)$ pattern, see Fig.\,\ref{Figure3} (b).

\subsection{Vortex glass: fracturing into
large non-hexatic domains}

Theoretical studies for order-disorder transitions in two dimensions
predict that between the ordered crystal and the fully disordered
liquid exists an intermediate hexatic phase presenting long-range
orientational and short-range positional
orders.~\cite{Kosterlitz1973} Whether this is a general scenario
that also holds for three-dimensional systems will provide useful
information for describing order-disorder transitions on general
grounds. A seminal work for the intermediate dimensionality case of
thick superconducting films reported a first-order transition to a
high-field vortex state with strongly reduced longitudinal
correlations, at odds with a hexatic vortex state.~\cite{Kes1986}

Figure\,\ref{Figure4} shows the evolution of the orientational order
at the  surface of the vortex structure (from MD data) on crossing
the Bragg to vortex glass transition. We characterized the
orientational order by means of the correlation function $G_{\rm
6}(r)= \langle \Psi_{6}(0) \Psi^{*}_{6}(r) \rangle$ that depends on
the distance-evolution of the hexagonal orientational order
parameter $ \Psi_{6} (\mathbf{r} = \mathbf{r_{i}}) =
\Sigma_{j=1}^{n}(1/n) \exp{(6i\theta_{ij})}$ calculated from the
bond angles of nearest
 neighbor vortices $i$ and $j$, $\theta_{ij}$. For $B/B_{\rm ord-dis} \leq 0.1$,
  $G_{\rm 6}$ starts decaying algebraically at
 short distances (see dashed lines) up to a characteristic length
 of the order of the size of the crystallites, see arrows. For larger
 distances the decay of $G_{\rm 6}$ is fitted by
 an exponential dependence. On further increasing field within the Bragg
 glass ($0.1 < B/B_{\rm ord-dis} < 1$), single crystalline vortex structures
  are observed and coincidentally $G_{\rm 6}$ decays algebraically at
all distances. The field-evolution of the exponent $\eta$ of this
algebraic decay is shown in Fig.\,\ref{Figure4} (c) for pristine and
electron-irradiated samples: it decreases dramatically when passing
from the polycrystalline to the single-crystalline structures and
remains almost constant for $B/B_{\rm ord-dis}>0.2$. This stagnation
is expected for a phase presenting long-range orientational order
such as the Bragg glass. Interestingly, the saturation value $\eta
\sim 0.025$ that we found at the surface of the vortex structure
nucleated in the Bragg glass phase is roughly one order of magnitude
smaller than the value found in the same phase from bulk SANS
measurements in single-crystalline
vanadium.~\cite{Toft-Petersen2018} This indicates that vortex
meandering within the sample thickness is significant even though
the orientational order of the Bragg glass is long-ranged in the
whole sample volume.

 For fields
above $B_{\rm ord-dis}$, the vortex glass phase presents a faster
decay of orientational order: even within the larger crystallite of
the center of  Fig.\,\ref{Figure2}, $G_{\rm 6}$ decays exponentially
with distance. Once the limits of the grain boundary are reached,
see black arrow in Fig.\,\ref{Figure4} (a), there is a kink in
$G_{\rm 6}$ and  the decay continues to be exponential. Therefore,
the orientational order of the vortex glass phase at the surface of
Bi$_{2}$Sr$_{2}$CaCu$_{2}$O$_{8 + \delta}$ samples is not
characterized by the typical algebraic decay of a hexatic phase.
This goes at odds with theories of the vortex glass being an hexatic
phase,~\cite{Chudnovsky} but is evocative of another theoretical
proposals of a multi-domain glassy phase separating the Bragg glass
and the vortex liquid.~\cite{Menon2002} The fracturing of the vortex
structure into large domains in the vortex glass phase could be a
non-equilibrium feature, due to finite cooling rates during the
nucleation of the structure in field-cooling
experiments.~\cite{Demirdis2016} Nevertheless, the vortex glass
presents non-hexatic order already inside the crystallites.

Data from SANS experiments probing  the structural properties of the
vortex glass in the whole thickness of the sample are consistent
with this degradation of orientational order  at the surface. The
intensity measured in a SANS experiment is the product of the
structure factor times the magnetic form factor $f(q)$ averaged over
the entire volume of the sample, $I(q)= S_{\rm
SANS}(q,\varPsi)\,\cdot\,f^{2}(q)$.~\cite{Pautrat2007} The structure
factor measured in SANS therefore collects information on the
meandering of vortices along the sample thickness, namely $S_{\rm
SANS}(q,\varPsi)= \mid \int \rho(q_{\rm x}, q_{\rm y},z) dz
\mid^{2}$. Figure \,\ref{Figure5} shows a comparison of the physical
magnitudes that can be accessed in SANS and MD experiments showing
data obtained in the Bragg and vortex glass phases of
Bi$_{2}$Sr$_{2}$CaCu$_{2}$O$_{8 + \delta}$.

Figure\,\ref{Figure6} shows the field-evolution of the SANS
intensity $I(q_{\rm Bragg})\,\cdot\, q_{\rm Bragg}$ measured at the
Bragg wave-vector $q_{\rm Bragg}=2\pi/a_{\rm 0}$ and  normalized by
its value at zero field, for pristine
Bi$_{2}$Sr$_{2}$CaCu$_{2}$O$_{8 + \delta}$. In the London limit when
vortices are sufficiently separated (see Supplementary material for
discussion on the validity of this limit), this magnitude is
field-independent only if $S_{\rm SANS}$ is constant. The figure
also includes previous data from (K, Ba)BiO$_{3}$, considered in the
literature as decisive to confirm experimentally that the phase at
$B < B_{\rm ord-dis}$ is the Bragg glass.~\cite{Klein2001} Similarly
as in our case, the authors of Ref.~\onlinecite{Klein2001}
considered $B_{\rm ord-dis} = B_{\rm ON}$, the onset field of the
second peak effect measured in (K, Ba)BiO$_{3}$. For $B/B_{\rm
ord-dis} < 1$, the data for both systems are remarkably similar
within the instrumental resolution of each experiment. For both
materials  the neutron intensity $I(q_{\rm Bragg})\,\cdot\, q_{\rm
Bragg}$
 is field-independent for $B
\lesssim 0.4 B_{\rm ord-dis}$ and decreases roughly exponentially
beyond $0.4 B_{\rm ord-dis}$. Therefore, our results in the
low-field phase of Bi$_{2}$Sr$_{2}$CaCu$_{2}$O$_{8 + \delta}$ are
consistent with the phenomenology observed in the Bragg glass phase
of (K, Ba)BiO$_{3}$.

In the vicinity of the glass-to-glass transition, the normalized
SANS intensity is of the order of 0.1 in both systems. For the
measurements in (K, Ba)BiO$_{3}$ of Ref.~\onlinecite{Klein2001} the
authors stated that the intensity was below the noise level for $B
\sim B_{\rm ord-dis}$ and beyond. In our measurements we are able to
detect a measurable intensity in the vortex glass phase since  we
have roughly a two orders of magnitude gain in the neutron flux with
respect to previous works, due to the virtuous combination of
measuring a larger sample and increasing significantly the counting
time. We are able to detect a non-negligible normalized neutron
intensity of $6 \times 10^{-2}$ well above the order-disorder
transition. This experimental resolution allowed us to explore the
structural properties of the vortex glass phase in
Bi$_{2}$Sr$_{2}$CaCu$_{2}$O$_{8 + \delta}$. We find that in the
vortex glass the neutron diffracted intensity also continues to
decay  with field, roughly exponentially, in almost two decades
more. For the field-range of our measurements in
Bi$_{2}$Sr$_{2}$CaCu$_{2}$O$_{8 + \delta}$, vortices are in the
London limit approximation and $f(q_{\rm Bragg})\,\cdot\, q_{\rm
Bragg}$ is constant with field, see discussion in Supplementary
material.\cite{Pautrat2007} Therefore the decay of $I(q_{\rm
Bragg})\,\cdot\, q_{\rm Bragg}$ in the vortex glass of
Bi$_{2}$Sr$_{2}$CaCu$_{2}$O$_{8 + \delta}$ comes from a reduction of
$S_{\rm SANS}$ with field. This implies that, undoubtedly, there is
a worsening of the structural properties in the vortex glass phase
of Bi$_{2}$Sr$_{2}$CaCu$_{2}$O$_{8 + \delta}$.

When measuring in Bragg condition in the plane of the detector, six
large diffraction peaks are always observed in the vortex glass
phase, even up to $B/B_{\rm ord-dis} = 1.8$. However, the average
azimuthal width of the peaks growths on increasing field. In order
to quantify the degradation of orientational order in the bulk of
the vortex structure,
 we measure the average azimuthal
width of the Bragg peaks $\sigma_{\perp}^{2}$ (in units of
$\AA^{-1}$), see the $I(\Psi,q_{\rm Bragg})$ profile in
Fig.\,\ref{Figure5} (b). Correcting  by the instrumental resolution
we  estimate the in-plane azimuthal correlation length
$\xi_{\perp}=1/(\sqrt{\sigma_{\perp}^{2} - \sigma_{\perp
inst}^{2}})$. This magnitude is associated with the typical distance
at which the shear displacements of vortices in the volume of the
sample are of the order of $a_{0}$.~\cite{Gammel1998}
Figure\,\ref{Figure7} (b) shows that the normalized
$\xi_{\perp}/a_{0}$ obtained from SANS experiments in sample P
decreases with field in the Bragg glass phase up to $B/B_{\rm
ord-dis}\sim 0.6$; then remains constant or slightly recovers
(difficult to ascertain within the experimental error) up to $B_{\rm
ord-dis}$, and finally systematically decreases with field in the
vortex glass. This comes from an azimuthal widening of the Bragg
peaks that is in agreement with MD evidence of a fracturing of the
vortex structure into large crystallites with small misalignment,
and non-hexatic orientational order inside them.

In order to quantitatively support this agreement, MD data can also
be analyzed in a similar fashion as to obtain the in-plane azimuthal
correlation length of the vortex structure at the sample surface.
The results after analyzing the azimuthal width of the Bragg peaks
observed in $S_{\rm MD}(q,\Psi)$, corrected by the MD instrumental
resolution (see details in the caption of Fig.\,\ref{Figure5}) are
shown with full circles in Fig.\,\ref{Figure7} (b). Turning on field
from zero to $B/B_{\rm ord-dis} \sim 0.1$, $\xi_{\perp}/a_{0}$
enhances for both pristine (black dots) and electron-irradiated
(blue dots) samples. For these low fields not covered by our SANS
experiments, the polycrystalline vortex matter observed by MD
presents larger crystallites on increasing vortex density. Then for
the range of fields were MD and SANS data are both available, the
perpendicular correlation length obtained in MD experiments
decreases and seems to stagnate with $B$ within the Bragg glass
phase. On the vortex glass phase, $\xi_{\perp}/a_{0}$ is slightly
smaller than in the Bragg glass. The absolute value of this
correlation length obtained in MD experiments is larger than in
SANS, and therefore the described evolution is pictorially less
evident in a semi-log plot as that of Fig.\,\ref{Figure7} (b). This
difference in absolute values can not be ascribed to the form factor
of vortices since this magnitude is roughly constant with field for
Bi$_{2}$Sr$_{2}$CaCu$_{2}$O$_{8 + \delta}$ at the studied field
range. Therefore, the difference comes from the inequality between
the structure factors probed by MD and SANS, the latter integrating
the meandering of vortices  along the sample thickness, the former
having information only from the location of vortices at the
surface. In addition, the SANS signal is collected in a larger
sample whereas MD data are obtained in one order of magnitude
smaller fields of view and the cumulative effect of topological
defects in decreasing the correlation lengths is smaller.

\subsection{Shortening of the positional order of the vortex glass}

Further quantitative characterization of the structural changes at
the glass-to-glass vortex transition can be gained by analyzing the
functional decay of the positional order as well as the radial
in-plane and longitudinal correlation lengths at the volume of the
sample.  Theoretically, elastic approaches showed that even though a
weak random disorder destroys the perfectly hexagonal vortex
structure, in the Bragg glass phase, quasi long-range in-plane
positional order  and algebraically divergent Bragg peaks in the
structure factor are expected in the volume of the
sample.~\cite{Giamarchi1995} Experimentally, previous data for
pristine samples show resolution-limited fine Bragg
peaks~\cite{Pautrat2007,Klein2001}, and images at the surface
reveals a distance-evolution of the positional correlation
function~\cite{Fasano2005} and displacement
correlator~\cite{Kim1995} consistent with the random manifold regime
of the Bragg glass. In this section we focus on the positional order
characteristic of the vortex glass phase hinging our analysis on
physical magnitudes that were also used to fingerprint the Bragg
glass phase.

In particular, a relevant magnitude to describe the positional
order is the displacement correlator, $W(r) = \langle [u(r) -
u(0)]^{2}\rangle/2$ with $u(r)$ the displacement of vortices with
respect to the sites of a perfect hexagonal lattice, and the average
taken over quenched disorder and thermal fluctuations. This
magnitude can in principle be obtained from MD data, but computing
$W(r)$ for structures with topological defects as observed
experimentally is not straightforward. Following a previous work of
some of us,  we implemented an algorithm to locally calculate $W(r)$
 in structures with defects.~\cite{Bolecek2017} The comparison is not done with a unique
perfect hexagonal structure in the whole field-of-view, but instead with
local perfect lattices with lanes oriented in the three principal directions
of the local structure. The regional lanes stop
running two lattice parameters away of any topological defect and
new lanes are defined if the structure slightly modifies its
orientation. Using this local algorithm, we obtain a modified average
displacement correlator  $W^{*}(r)$ for structures with topological defects.
The Supplementary material section includes a schematics on
 how this analysis is performed.

Figure\,\ref{Figure4}\,(b) presents the distance-evolution of
$W^{*}/a_{0}^{2}$ for various $B/B_{\rm ord-dis}$ ratios on
traversing the order-disorder transition for  electron irradiated
samples A and B. The main panel of Fig.\,\ref{Figure4} (b) shows
that for $B/B_{\rm ord-dis}<1$, $W^{*}/a_{0}^{2} \propto
(r/a_{0})^{\nu}$, see fits presented with dashed lines. The same
phenomenology is observed in the Bragg glass phase of P samples
studied here, see the Supplementary material section. The exponent
$\nu$ for both electron-irradiated an pristine samples decays with
$B/B_{\rm ord-dis}$ in the Bragg glass phase, see
Fig.\,\ref{Figure4}\,(c). For P samples $\nu$ stagnates for
$B/B_{\rm ord-dis} > 0.2$ around the value of 0.44  indicated with a
dotted line in the figure. The algebraic decay of $W^{*}/a_{0}^{2}$,
as well as this value of $\nu$, are theoretically expected for the
random-manifold regime of the Bragg glass.~\cite{Giamarchi2002} Even
for the case of the polycrystalline vortex structures observed at
$B/B_{\rm ord-dis} \leq 0.1$, the growth of $W^{*}/a_{0}^{2}$
follows the same functionality inside the crystallites (distances
below the arrow in Fig.\,\ref{Figure4} (b)). In contrast, for
electron irradiated samples, $\nu$ does not seem to stagnate for the
studied fields, but reaches a value of $\sim 0.5$ for $B/B_{\rm
ord-dis} = 0.75$. These values of $\nu$ for the Bragg glass phase
are roughly one order of magnitude larger than those found for
describing the decay of orientational order, $\eta$, indicating bond
orientational order in the Bragg glass phase is of longer range, as
compared to translational order.

For the vortex glass phase, the growth of $W^{*}/a_{0}^{2}$ can
still be fitted with an algebraic growth, but the exponent found,
$\nu = 0.7$, is significantly larger than in the Bragg glass. This
behavior is found already at distances smaller than the typical
crystallite size in the vortex glass phase, see data in the  insert
to Fig.\,\ref{Figure4} (b) for $B/B_{\rm ord-dis}=1.62$ with a
crystallite size of $\sim 20a_{0}$.  In addition, also the absolute
value of the cumulated displacements at a fixed distance changes on
traversing the order-disorder transition. For instance,
Fig.\,\ref{Figure4} (d) shows the evolution of $W^{*}/a_{0}^{2}$ at
$r/a_{0}=10$ with increasing field: in the Bragg glass phase
decreases up to $B/B_{\rm ord-dis}=0.2$, roughly stagnates up to the
order-disorder transition, and has a significantly larger value in
the vortex glass phase.   All this evidence suggests that the vortex
glass presents short-range positional order with $u(r)$ growing
faster  than in the Bragg glass phase.

We  further characterized the nature of the positional order of the
vortex glass by studying the correlation lengths in the volume of
the sample from SANS data. The first magnitude that we study is the
radial in-plane correlation length $\xi_{\parallel}$, associated
with compressive displacements of the vortex structure convoluted
all along the thickness of the sample. This magnitude is estimated
from the radial width of the Bragg peaks minus the instrumental
resolution,  namely $\xi_{\parallel} = 1/
\sqrt{\sigma_{\parallel}^{2}- \sigma_{\parallel inst}^{2}}$, see
 Fig.\,\ref{Figure5} (b). The average $q$-width of the Bragg peaks,
 $\sigma_{\parallel}$, is only slightly larger
than the experimental resolution in the Bragg glass, but for
$B>B_{\rm ord-dis}$  dramatically increases with field far beyond
the instrumental resolution, see Fig.\,\ref{Figure6} (b). First,
this implies that diffraction peaks in the vortex glass phase are
not resolution-limited as in the Bragg glass, but widen beyond
resolution. Second, the drastic widening of the peaks yields a
systematic reduction of the in-plane radial correlation length with
field, even in units of $a_{0}$, as shown in Fig.\,\ref{Figure7}
(a). The data for sample P obtained with SANS are consistent with
the reduction of $\xi_{\parallel}$ in units of $a_{0}$ also observed
in MD data at the surface of pristine and electron-irradiated
samples. Similarly as in the case of the azimuthal in-plane
correlation length, MD data are generally above the SANS data since
the latter convolute information in the volume of the sample.
Comparative examples of the radial width of Bragg peaks in MD and
SANS data, indicating the instrumental resolution in both
experiments, are shown in the Supplementary material.

We get extra information on the structural properties along the
direction of vortices by measuring SANS rocking curves tilting the
sample around the $\omega$ direction in the Bragg condition, see
Fig.\,\ref{Figure5} (b). The width of the rocking curve, once
corrected by the instrumental resolution, provides an estimation of
the longitudinal correlation length associated with tilting
displacements, $\xi_{\rm L} = 1/ \sqrt{\sigma_{\omega}^{2}-
\sigma_{\omega\,\,inst}^{2}}$. This characteristic length measures
the correlation along flux lines and is reached at a $z-$position
where the lateral displacement of a vortex with respect to a
straight line is of the order of coherence length. Previous results
in (K,Ba)BiO$_3$ indicated that in the Bragg glass phase the width
of the rocking curve is resolution-limited (remains roughly
constant), at least within an experimental resolution of $\Delta
\omega \sim 0.18$\, degrees.~\cite{Klein2001}

In our measurements in Bi$_{2}$Sr$_{2}$CaCu$_{2}$O$_{8 + \delta}$,
with an experimental resolution of $\Delta \omega \sim
0.1$\,degrees, the corrected width of the rocking curve is not
changing appreciably in the Bragg glass up to $B/B_{\rm ord-dis}
\sim 0.6$. However, very close to the order-disorder transition, at
$B/B_{\rm ord-dis} \sim 0.8$, the rocking curve shrinks and
$\xi_{\rm L}/a_{0}$ increases around 30\,\%. On entering the vortex
glass phase, the longitudinal correlation length seems to smoothly
decrease, although error bars are quite large since the diffracted
intensity decreases exponentially with field. Nevertheless, we have
sufficient instrumental resolution as to ascertain that no dramatic
collapse of $\xi_{\rm L}/a_{0}$ is observed in the vortex glass
phase. The shortening of the positional order in the vortex glass is
more dramatic for in-plane displacements than for longitudinal ones.
The latter does not rule out the possibility that some screw
dislocations could nucleate along the direction of vortices in the
vortex glass. These are very relevant findings since they suggest
that the long-range order of the superconducting phase is maintained
in the vortex glass. Indeed, Josephson plasma resonance data show
that in the vortex glass the c-axis correlations are enhanced with
respect to the vortex liquid.~\cite{Gaifullin2000} As a consequence,
our results imply, as proposed by other
groups,~\cite{Safar1992,vanderBeek1992,Beidenkopf2005} that
superconducting phase coherence should be destroyed in a transition
line between the vortex glass and the vortex liquid.

In an elastic description of the vortex lattice the ratio between
the longitudinal and in-plane correlation lengths are
proportional to the tilting energy, namely $\xi_{\rm L}/\xi_{\perp,
\parallel} \sim
\sqrt{c_{44}/c_{66}}$.~\cite{Blatter}  To evaluate this ratio in the case of
Bi$_{2}$Sr$_{2}$CaCu$_{2}$O$_{8 + \delta}$ with $\lambda=2$\,$10^{-5}$\,cm, we first
consider that $c_{44} = B^{2}/4\pi(1 + \lambda^{2}k^{2})
\approx B^{2}/4\pi$ since we are taking into account elastic distortions with $k
\sim 1/\xi_{L} \sim 10^{-4}$\,cm, and then
 $\lambda^{2}k^{2} \sim 10^{-2} \ll 1$. Second, considering the value of $\lambda$ for
 this material and $\Phi_{0} = 2.07$\,$10^{-7}$\,G.cm$^2$ the shear elastic constant
 $c_{66} \approx \Phi_{0}B/(8\pi \lambda)^{2} \sim
B$. These considerations yield $\sqrt{c_{44}/c_{66}} \simeq
\sqrt{B/4\pi}$ in an elastic description for
Bi$_{2}$Sr$_{2}$CaCu$_{2}$O$_{8 + \delta}$. The insert to
Fig.\,\ref{Figure7} (c) shows the field-evolution of the
longitudinal (and radial) to azimuthal correlation lengths as
measured by SANS. Within the error bars, these ratios are very close
to $\simeq  \sqrt{B/4\pi}$. This suggests that the eventual density
of screw dislocations is not as large as to plastic deformations
along the direction of vortices being dominant in the vortex glass.

\section{Discussion}

The decrease of the three correlation lengths (not so dramatic for
$\xi_{\rm L}$) on transiting  in the vortex glass, detected both via
MD and SANS, is accompanied by a reduction of the diffracted
intensity $I(q_{\rm Bragg}).q_{\rm Bragg}$. This intensity  starts
decaying for $B/B_{\rm ord-dis} \sim 0.4$ and close to the
order-disorder transition registers one decade of decrease. Although
this quantitative evolution was also found in SANS measurements in
(K,Ba)BiO$_3$ and and in (Ba,K)Fe$_{2}$As$_{2}$, in our SANS
experiments in Bi$_{2}$Sr$_{2}$CaCu$_{2}$O$_{8 + \delta}$, we still
have a good instrumental resolution for $B
> B_{\rm ord-dis}$. In our case, this collapse  is mainly due to a falling
down of the structure factor
$S_{\rm SANS}(q,\Psi)$ since for the studied fields the form factor
is roughly field-independent. The concomitant widening of
the Bragg peaks and collapse of diffracted intensity measured by SANS
at the volume are consistent with
 the non-hexatic vortex glass that fractures into large
crystallites with small misalignment and presents a shortening of
positional order as revealed at the surface by MD. These results in
Bi$_{2}$Sr$_{2}$CaCu$_{2}$O$_{8 + \delta}$ are consistent with SANS
data in the pnictide material
(Ba,K)Fe$_{2}$As$_{2}$.~\cite{Demirdis2016} Therefore, in this
particular order-disorder transition no two-stage structural
transformation is detected.

The vortex system in this extremely layered superconductor undergoes
a first-order glass-to-glass transition between a quasi-crystalline
Bragg glass to a non-hexatic and short-range positionally ordered
vortex glass with no dramatic loss of correlation along the
direction of vortices.  Strikingly, we obtain our results in an
extremely layered type-II superconductor with flux lines formed by
columns of two-dimensional pancake vortices that are longitudinally
weakly-bounded, specially at high temperatures. The slight decay of
$\xi_{\rm L}/a_{0}$, remaining non-negligible for $B/B_{\rm ord-dis}
> 1$, rules out the possibility of the vortex glass being a phase of
decoupled pancake vortices along the field direction. Moreover,
since the ratio between the longitudinal and in-plane correlation
lengths follows quite well the prediction for an elastic description
of the vortex glass, the proliferation of screw dislocations along
the direction of vortices seems to be unlikely. Therefore, our
evidence suggests that in the vortex glass phase of
Bi$_{2}$Sr$_{2}$CaCu$_{2}$O$_{8 + \delta}$, even though this
material is extremely layered,  the flux lines behave as
three-dimensional, formed by coupled pancake vortices along the
c-axis direction. This might be at the origin of the glass-to-glass
transition in Bi$_{2}$Sr$_{2}$CaCu$_{2}$O$_{8 + \delta}$ not being
well described by the two-stage melting
description.~\cite{Kosterlitz1973} Therefore our results reminds
that ubiquitous order-disorder transitions in nature, either melting
or glass-to-glass, are richer than the scenario predicted by this
theory for melting in two-dimensions. In particular, the
dimensionality of vortex matter controlled by the electronic
anisotropy of the host superconductor plays a determinant role in
the structural properties of the disordered vortex glass phase.

Even though our results are a comprehensive study of the structural
properties of the vortex glass in the surface and volume of the
vortex ensemble, we surveyed these properties in a static condition
after field-cooling.  However, we have to have always in mind that
structural transitions in vortex matter, and in systems with
substrate disorder in general, are actually transitions in a state
of matter different from a solid or a liquid, particularly regarding
their dynamical properties. Experiments probing the dynamics of both
glassy phases are thus important to further confirm our findings.

\section*{Methods}

\textbf{Sample preparation}

We studied electron-irradiated as well as  pristine
Bi$_{2}$Sr$_{2}$CaCu$_{2}$O$_{8 + \delta}$ samples from different
sample growers. Pristine optimally-doped samples used in MD
experiments (roughly 30 small single crystals) were grown, annealed
and characterized at both, the Low Temperature Lab of Bariloche,
Argentina, and the Kamerlingh Onnes Lab at Leiden, The Netherlands.
SANS experiments  were performed using a large single crystal of $30
\times 5 \times 1.2$\,mm$^{3}$ grown at the International
Superconductivity Technology Center of Tokyo, Japan, by A. Rykov,
and characterized at Ensicaen, France.
 The slightly-overdoped electron-irradiated samples (grown in
a 200\,mbar O$_2$ atmosphere at Leiden) were irradiated with
2.3\,MeV electrons at low temperatures (20\,K) in a van de Graaff
accelerator coupled to a closed-cycle hydrogen liquifier at the
\'{E}cole Polytechnique of Palaiseau,
France.~\cite{Konczykowski2009} Two single crystals with a
significant decrease of the $B_{\rm ord-dis}$ order-disorder
transition field were selected for the MD study. Sample A was
annealed at 793\,C in air and then irradiated with a dose of
$1.7\times 10^{19}$\,e/cm$^2$; sample B was not annealed and
irradiated with $7.4\times 10^{19}$\,e/cm$^2$. Electrons traverse
the whole sample thickness generating a roughly homogeneous
distribution of point defects in the crystal structure. Irradiating
at temperatures lower than the threshold temperature for defect
migration is essential in order to prevent the agglomeration of
defects. Some of the point defects annihilate on warming the
samples, but the remaining defects are of atomic size
.~\cite{Konczykowski2009}

\textbf{Local Hall magnetometry}

We applied local Hall probe magnetometry in order to track the
changes produced in the vortex phase diagram by the point-defect
potential introduced by electron irradiation. The local stray field
of the samples is measured with an array of  GaAs/AlGaAs Hall probes
with active areas of 16\,$\times$16\,$\mu$m$^{2}$.~\cite{Dolz2014}
We performed dc and ac measurements applying constant, $H$, and
ripple, $h_{\rm ac}$, fields parallel to the c-axis of the samples.
Measuring the sample magnetization, $H_{\rm s} = B - H$, when
cycling $H$ at fixed temperatures allow us to obtain dc magnetic
hysteresis loops as shown in Fig.\,\ref{FigureSM1} (a). ac
transmittivity measurements are performed by simultaneously
acquiring the first and third harmonics of the ac magnetic induction
when applying a ripple $h_{\rm ac}$ field either by changing
temperature at fixed $H$ or cycling $H$ at fixed temperature. The
transmittivity $T'$ is obtained by normalizing the in-phase
component of the first-harmonic signal $B'$, namely $T'=[B'(T) -
B'(T \ll T_{\rm c})]/[B'(T>T_{\rm c}) - B'(T \ll T_{\rm
c})]$.~\cite{Gilchrist1993} This magnitude is extremely sensitive to
discontinuities in the local induction as the $B$-jump produced at
the first-order vortex transition at $B_{\rm FOT}$. The third
harmonic signal, $\mid T_{\rm h3} \mid= \mid B_{h3}^{\rm AC}
\mid/[B'(T>T_{\rm c}) - B'(T \ll T_{\rm c})]$, is measured to detect
the onset of non-linearities in the magnetic response, see
Fig.\,\ref{FigureSM1} (d). ac measurements were typically performed
with ripple fields of 1\,Oe  amplitude and  7.1\,Hz frequency.
Further details are discussed in the Supplementary material section.

\textbf{Magnetic decoration}

In order to directly image individual vortex positions in a typical
field-of-view of 1000-5000 vortices, MD experiments were performed
simultaneously in electron-irradiated and  pristine
Bi$_{2}$Sr$_{2}$CaCu$_{2}$O$_{8 + \delta}$ samples. The structural
properties of vortex matter were imaged below and above the $B_{\rm
SP}$ order-disorder transition by field-cooling the samples  from
room temperature down to 4.2\,K at applied fields  $5<H<150$\,Oe.
Further details in the decoration protocol followed in this case can
be found in Ref.\,\cite{Fasano1999}. Magnetic decorations were
performed at different fields in roughly 30 pristine freshly-cleaved
small single crystals from two sample growers. For every field,
experiments were performed at several realizations for statistical
purposes. In the case of the two electron-irradiated samples,
experiments were performed at different fields on subsequently
cleaving the samples. Since this is a destructive process, every
sample was first studied with Hall magnetometry and then decorated.

\textbf{Small angle neutron scattering}

SANS experiments were performed at the D22 diffractometer of the
Institut von Laue Langevin at Grenoble, France. The wavelength of
incident neutrons was of 9 and 15\,\AA\ with a wavelength resolution
of 10\,\%. The collimation of the incident beam produced a beam
divergence of $\delta(2\theta)_{div}\sim 0.1 \times 10^{-3}$\,rad.
The angular distribution of the scattered intensity, $I(q,\Psi)$,
was measured in a $102.4 \times 98$\,cm$^2$ detector with $1280
\times 1225$\,pixel$^2$ ($0.8 \times 0.8$\,mm$^2$ per pixel) located
at 17.6\,m from the sample. SANS data were obtained by rocking
horizontally ($\phi$ direction) and vertically ($\omega$ direction)
the whole sample and magnet system when aligned in the Bragg
condition.~\cite{Pautrat2007} In order to obtain the signal coming
purely from the vortex lattice, we subtracted from the raw data the
normal-state background signal measured at zero field and 10\,K.
Similarly as in MD experiments, vortex diffraction patterns and
rocking curves were measured at 4.2\,K after field-cooling the
sample in a magnetic field applied along the c-axis and with
magnitude ranging 100 to 1000\,Oe.

\section{Supplementary material}

\subsection{First-order transition and irreversibility line from ac
and dc Hall magnetometry }

 The first-order high-temperature $B_{\rm FOT}$ and
order-disorder $B_{\rm ord-dis}$ transitions, as well as the
irreversibility line of vortex matter for the studied samples were
obtained combining ac and dc local magnetometry measurements.
Figures \,\ref{FigureSM1} (a) to (d) show these results in the
illustrative case of sample B. The glass-to-glass transition  is
detected from the onset of the local $H_{\rm S}$-peaks in the
ascending and descending branches of dc magnetization loops measured
at low temperatures, see Fig.\,\ref{FigureSM1} (a). These peaks
start to develop at $B_{\rm ON}$ considered as the order-disorder
transition field $B_{\rm ord-dis}$; the local maxima in
magnetization  occurs at the field indicated as $B_{\rm SP}$ that
unlike $B_{\rm ON}$ depends on the field sweep-rate. Assuming a Bean
field-profile inside the sample, the critical current at a given
temperature can be estimated as proportional to the separation
between the ascending and descending branches of the magnetization
loop. The result of this analysis is shown in Fig.\,\ref{FigureSM1}
(b): the $B_{\rm ON}$ field coincides with the start of the $J_{\rm
c}$ peak and at $B_{\rm SP}$ the local maximum in critical current
is developed. This phenomenology for the detection of $B_{\rm
ord-dis}$ is consistent with the one followed in the literature for
pristine as well as electron-irradiated
Bi$_{2}$Sr$_{2}$CaCu$_{2}$O$_{8 + \delta}$
samples.~\cite{Khaykovich97a,Dolz2014}

The glass-to-glass $B_{\rm ord-dis}$ line is the continuation of the
high-temperature solid-to-liquid vortex transition $H_{\rm FOT}$,
~\cite{Avraham2001} and since both are of first order, a jump on $B$
is observed when controlling $H$.  Since this transition is of
first-order, a jump in $B$ is observed when controlling $H$. ac
transmittivity is a normalized derivative of the in-phase component
of the first harmonic of $B$. Therefore, at high temperatures the
transition is observed as a paramagnetic peak in
$T'(B)$,~\cite{Dolz2014} as for instance shown in
Fig.\,\ref{FigureSM1} (c).  In all the studied samples, the
field-position of the paramagnetic peak is independent of the
frequency and amplitude of $h_{\rm ac}$. For the studied samples,
the slope of the $B_{\rm FOT}(T)$ line is not dramatically altered
by electron irradiation, but since the $T_{\rm c}$ is significantly
affected the vortex solid phase spans a smaller region on increasing
the irradiation dose.

We also studied the  irreversibility line, a crossover line at which
pinning sets in and the magnetic response becomes therefore
irreversible. This is also observed as a non-linear magnetic
response, namely as a non-negligible value of the higher harmonics
of the ac magnetization signal. We therefore followed the standard
procedure of measuring the phase location of $B_{\rm IL}$ from the
onset of the third harmonic signal, $\mid T_{\rm h3}
\mid$.~\cite{Dolz2014} Figure\,\ref{FigureSM1}\,(d) shows that on
decreasing temperature $\mid T_{\rm h3} \mid$ overcomes the typical
noise level at a characteristic temperature indicated as $T_{\rm
IL}$ and then a peak-like structure develops on further cooling.
These data are shown for the illustrative case of sample B but
sample A and the pristine samples follow the same
phenomenology.~\cite{Dolz2014} The irreversibility line is not a
transition but a pinning crossover line and therefore strongly
depends on the amplitude and frequency of the ripple field applied
to measure ac magnetization.~\cite{Dolz2014} The data shown in
Fig.\,\ref{FigureSM1}\,(d) were obtained for an $h_{\rm ac}$ of 1\,Oe
and 7.1\,Hz.

\subsection{SANS intensity as a function of field: validity of the
London approximation for the form factor}

The decrease of $I(q_{\rm Bragg}).q_{\rm Bragg}$ for fields larger
than $B/B_{\rm ord-dis}=0.4$ was proposed in
Ref.~\onlinecite{Klein2001} as due to the proliferation of
topological defects on approaching $B_{\rm ord-dis}$. Our MD and
SANS results in Bi$_{2}$Sr$_{2}$CaCu$_{2}$O$_{8 + \delta}$
considered together, indeed show that there is a worsening of the
structural properties of vortex matter on approaching $B_{\rm
ord-dis}$, and that the in-plane and longitudinal correlation
lengths become significantly smaller in the vortex-glass phase,
surely associated with the observed proliferation of defects in MD
images. Nevertheless, the significant decrease of the neutron signal
in the case of (K, Ba)BiO$_{3}$ can be mainly associated with a
decrease of the magnetic form factor. Even for a perfect lattice in
which the structure factor is equal to 1, the magnitude $I(q_{\rm
Bragg}.q_{\rm Bragg}$ can be considered constant only in the London
limit in which vortices are far apart. However, on increasing field
close to $H_{\rm c2}$ the London limit is no longer valid and the
measured intensity for a perfect lattice decreases just because the
magnetic form factor of the vortex structure decreases with field.
For instance, and for the case of large $\kappa$ materials as
studied here, Brandt showed in Ref.\,\onlinecite{Brandt1997} that
the neutron diffracted intensity in a Ginzburg-Landau model
significantly decreases with field even for $B \sim 10^{-2} B_{\rm
c2}$.

Considering the data of Table 1 of Ref.\,\onlinecite{Brandt1997},
and the London expression for the neutron diffracted intensity, we
plot in Fig.\,\ref{FigureSM4} the ratio between the neutron
intensity in a Ginzburg-Landau to a London model. We have also added
gray regions indicating the range of fields at which SANS
measurements were performed for the two discussed type-II
superconductors. Clearly our measurements in
Bi$_{2}$Sr$_{2}$CaCu$_{2}$O$_{8 + \delta}$ lie in the $B/B_{\rm
c2}$ range in which the neutron diffracted intensity can be
reasonably well approximated by the London model, and any decrease
of intensity can only have origin in a worsening of the structure
factor. However, in the case of the field-range of the SANS
measurements of Ref.\,\onlinecite{Klein2001} in (K,Ba)BiO$_{3}$, the
reduction in the neutron diffracted intensity can also have origin
in the decrease of the magnetic form factor with field, for the field
range studied in that work.

\subsection{MD structure factor at the surface
and SANS intensity in the bulk of vortex matter}

Figure\,\ref{FigureSM5} shows $q$-profiles of the Bragg peaks as
detected from the structure factor measured by MD (full symbols) at
the sample surface and the SANS intensity (open symbols) convoluting
information along the whole sample thickness. Data in the left panel
shows the case of the Bragg glass phase whereas data at the vortex
glass phase are presented in the right panel. The gray boxes
highlight the experimental resolution for SANS and the dark gray
ones for MD experiments. The SANS instrumental resolution is given mainly by the dispersion in
the wavelength of neutrons and the divergence of the neutron beam on the detector.
The MD
resolution is estimated from considering the error in digitalizing
vortex positions (depending on the number of pixels per microns of a
given image), plus the typical width of the Bragg peak of a perfect
lattice with the same number of vortices in a region imaged with the
same amount of pixels than in the experiment.

\subsection{The lanes algorithm to calculate $W*$ in a structure
with topological defects}

The positional correlator was calculated using the algorithm of
lanes that allow us to calculate $W^{*}(r)$ in the case of lattices
with topological defects. The classical way to calculate the
displacement correlator is to measure the transversal, $u_{\rm t}$,
and longitudinal, $u_{\rm l}$, displacements of a vortex with
respect to the nearest site of a perfect hexagonal structure
oriented in a fixed direction in the whole image. This direction is
obtained from the location of the peaks in the Fourier transform of
the experimental structure. However, the presence of topological
defects locally alters the orientational order of the structure, and
if defects proliferate, the three vortex lanes of the structure can
significantly change their orientation. In this case, comparing with
the sites of a perfect structure with fixed orientation is not a
proper way of calculating the displacement correlator since
orientational information is entangled with the positional one.

We solved this problem by considering a local perfect lattice whose
three lanes are defined from the local orientation of the
experimental structure.  Therefore the orientation of the three
lanes of the perfect structure, $\mathbf{a}_{\rm i}$ with $i=0, 1,
2$ changes when a topological defect is reached. Once these lanes
are locally defined, the longitudinal $u^{i}_{\rm L}$ and
transversal $u^{i}_{\rm T}$ displacements are measured to obtain
$W^{i}_{T}(r) = \langle [u^{i}_{T}(r) - u^{i}_{T}(0)]^{2} \rangle$
and $W^{i}_{L}(r) = \langle [u^{i}_{L}(r) - u^{i}_{L}(0)]^{2}
\rangle$. Then $W_{T}(r)$ and $W_{L}(r)$  are obtained by averaging
the $W^{i}_{T}(r)$ and $W^{i}_{L}(r)$ measured with respect to every
lane direction $\mathbf{a}_{\rm i}$. Finally, the average
displacement correlator is computed as $W^{*} = (W_{T} + W_{L})/2$.
An example of this analysis with the local lanes is shown in
Fig.\,\ref{FigureSM2}. The results are shown with error bars
corresponding to the dispersion in $W^{*}$ values for every
$r/a_{0}$.

\subsection{Orientational and positional order of the Bragg glass in
pristine Bi$_{2}$Sr$_{2}$CaCu$_{2}$O$_{8 + \delta}$}

The positional and orientational order of the Bragg and vortex-glass
phases was characterized in real space by calculating the
displacement correlator $W^{*}$ and the orientational correlation
function $G_{\rm 6}$ from the vortex positions observed by MD. In
the main text data for the electron-irradiated samples A and B are
shown. Here we show also the data for the pristine samples P for the
range of fields $B/B_{\rm ord-dis}<1$ in which we have access to the
structure with single vortex resolution by applying the MD
technique. Results are consistent with data in the Bragg glass phase
of electron irradiated samples: the orientational correlation
function $G_{6}$ decays algebraically at intermediate and long-range
and the displacement correlator $W^{*}(r)$ grows algebraically with
distance as expected in the random manifold regime of the Bragg
glass. The exponents of both algebraic evolutions with distance are
shown in Fig.\,\ref{Figure4} (c) of the main text.

\section{Acknowledgments} This work was supported by the ECOS-Sud-MINCyt
France-Argentina bilateral program under Grant A09E03; by the
Argentinean National Science Foundation (ANPCyT) under Grants
PRH-PICT 2008-294 and PICT 2011-1537; by the Universidad Nacional de
Cuyo research grants program; and by Graduate Research fellowships
from IB-CNEA for J. A. S and from Conicet for R. C. M., G. R. and N.
R. C.B. We thank to M. Li and A. Rykov for growing some of the
studied pristine single crystals and V. Moser for providing us the
Hall sensors.

\section{Author contributions}
Y. F., A. P., M. K., and C. J. vd B. designed research;
all authors performed research; J. A. S., Y. F. and A. P. analyzed
data; Y. F. wrote the paper with the collaboration of the rest of
the authors.
The authors declare no conflict of interest.\\
\textsuperscript{*}To whom correspondence
should be addressed. E-mail: yanina.fasano@cab.cnea.gov.ar

\begin{figure*}[ttt]
\centering
\includegraphics[width=0.9\textwidth]{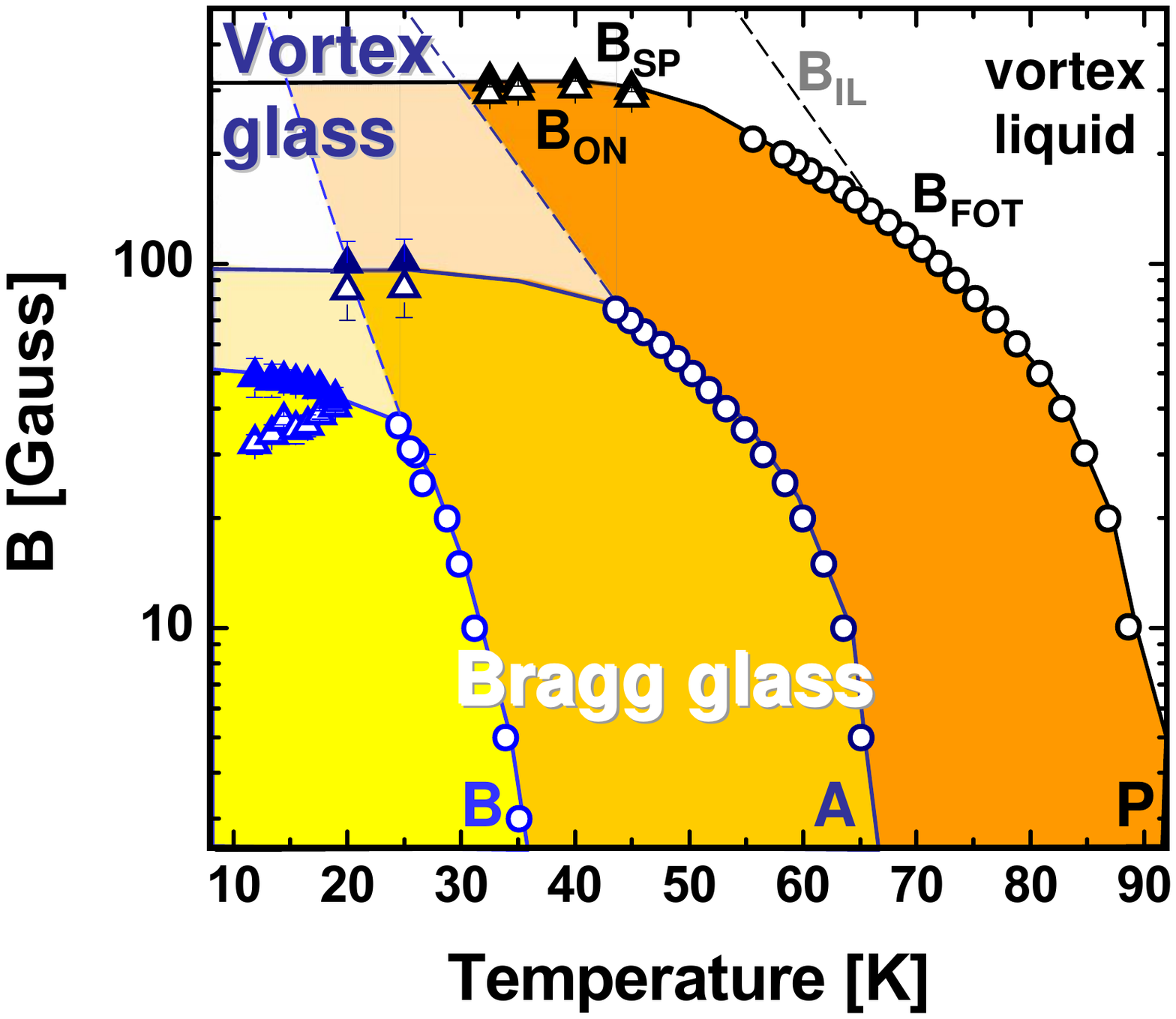}
\caption{Vortex phase diagram for pristine (P) and
electron-irradiated (A and B) Bi$_{2}$Sr$_{2}$CaCu$_{2}$O$_{8 +
\delta}$ samples  from local dc and ac Hall magnetometry
measurements. Sample A was irradiated with $1.7\cdot
10^{19}$\,e/cm$^2$ and sample B with $7.4\cdot 10^{19}$\,e/cm$^2$.
First-order transition lines $B_{\rm FOT}$ (open circles) and
$B_{\rm ord-dis}=B_{\rm ON}$ (open triangles) separate the Bragg
glass from the vortex liquid phase at high temperatures, and the
vortex glass at high fields. Full triangles signpost the location of
the so-called second peak in the critical current $B_{\rm SP}$.
Dashed lines represent the irreversibility line $B_{\rm IL}$,
located very close to $B_{\rm FOT}$ for the fields at which the
Bragg glass is stable. Solid lines are guides to the eye.}
\label{Figure1}
\end{figure*}

\begin{figure*}[hhh]
\includegraphics[width=0.9\textwidth]{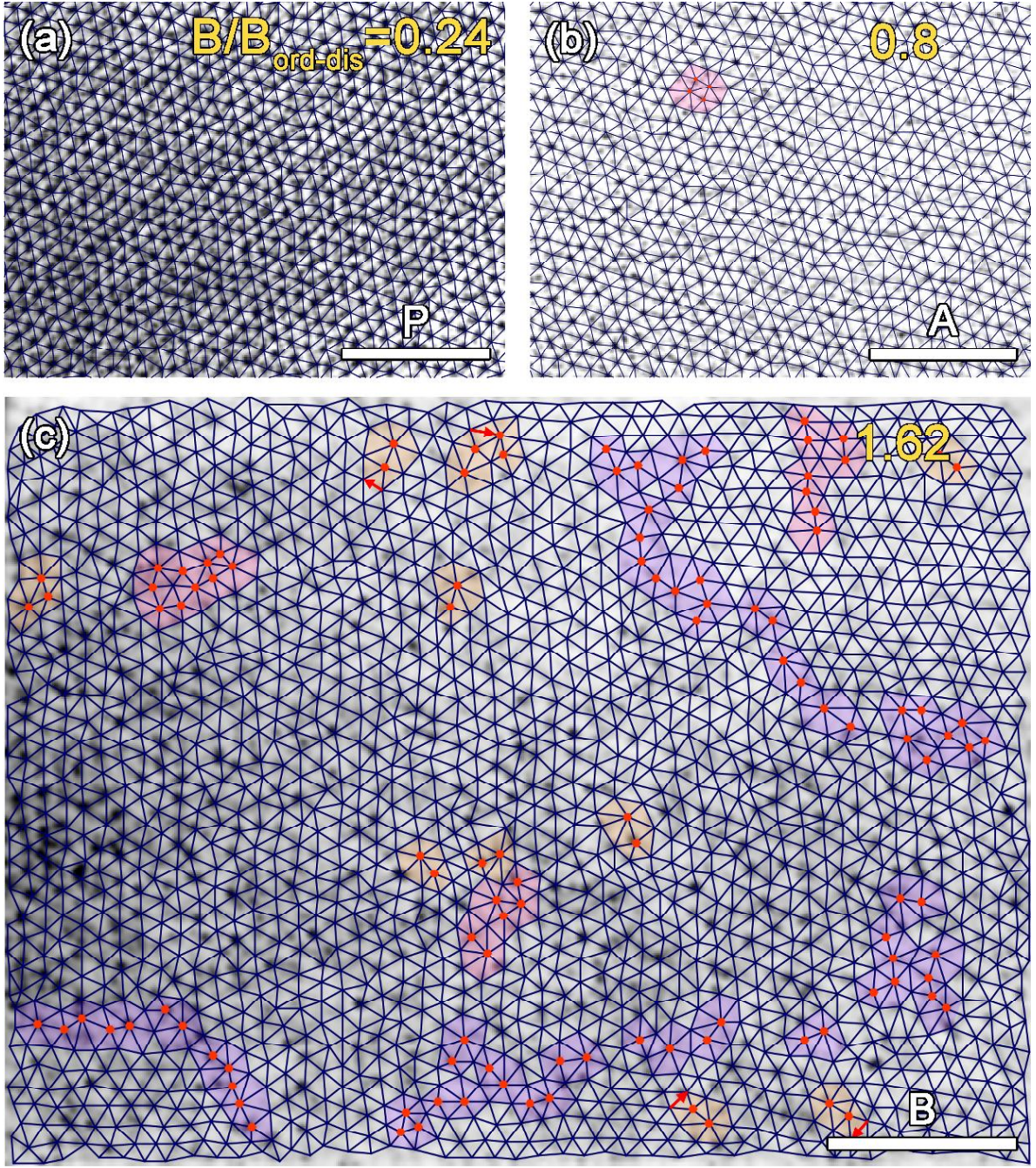}
\caption{Magnetic decoration images of individual vortices (black
dots) in the Bragg and the vortex glass phases of pristine (P) and
electron-irradiated (A and B) Bi$_{2}$Sr$_{2}$CaCu$_{2}$O$_{8 +
\delta}$ samples. Vortex structures in the Bragg glass phase for (a)
$B/B_{\rm ord-dis}=0.28$ in sample P and (b) $B/B_{\rm ord-dis} =
0.8$ in sample A. (c) Large field-of-view snapshot of the
vortex-glass phase observed in sample B for $B/B_{\rm
ord-dis}=1.62$. Delaunay triangulations are superimposed on the
structures: neighboring vortices are connected by dark blue lines
and non-sixfold coordinated vortices are highlighted in red.
Topological defects indicated with colors: grain boundaries
highlighted in violet, edge dislocations in orange and twisted bonds
in pink. Red arrows indicate the Burgers vectors of paired edge
dislocations; no arrows are shown if dislocations seem to be
unpaired in our experimental field-of-view. White bars indicate
5\,$\mu$m.} \label{Figure2}
\end{figure*}

\begin{figure*}[ttt]
\includegraphics[width=0.9\textwidth]{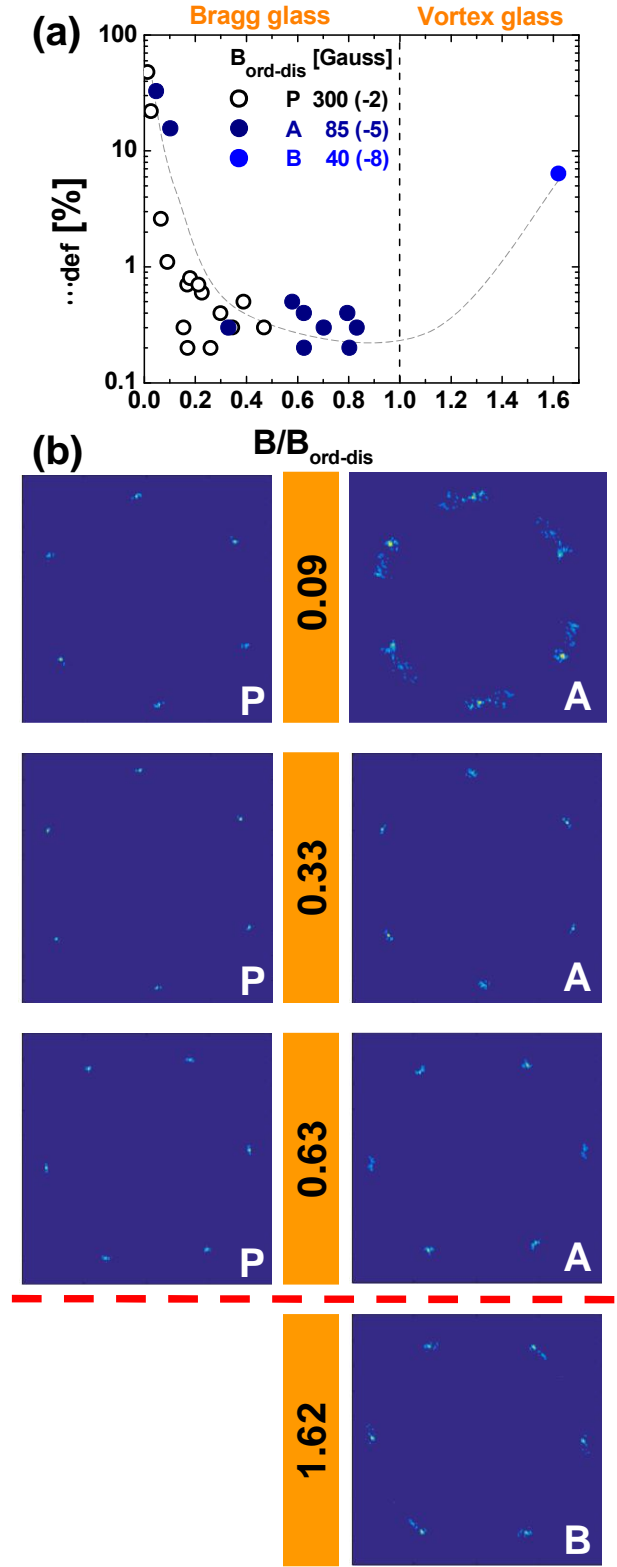}
\caption{Topological defects and structure factor data for the Bragg
and vortex glass phases of Bi$_{2}$Sr$_{2}$CaCu$_{2}$O$_{8 +
\delta}$ vortex matter at the surface of the samples. (a) Fraction
of non-sixfold coordinated vortices, $\rho_{\rm def}$, as a function
of $B/B_{\rm ord-dis}$. (b) Structure factor $S_{\rm MD}(q,\Psi)$ of
the vortex structure observed  by magnetically decorating the
individual vortex positions in pristine P (left panels) and
electron-irradiated A and B (right panels) samples.} \label{Figure3}
\end{figure*}

\begin{figure*}[ttt]
\includegraphics[width=0.9\textwidth]{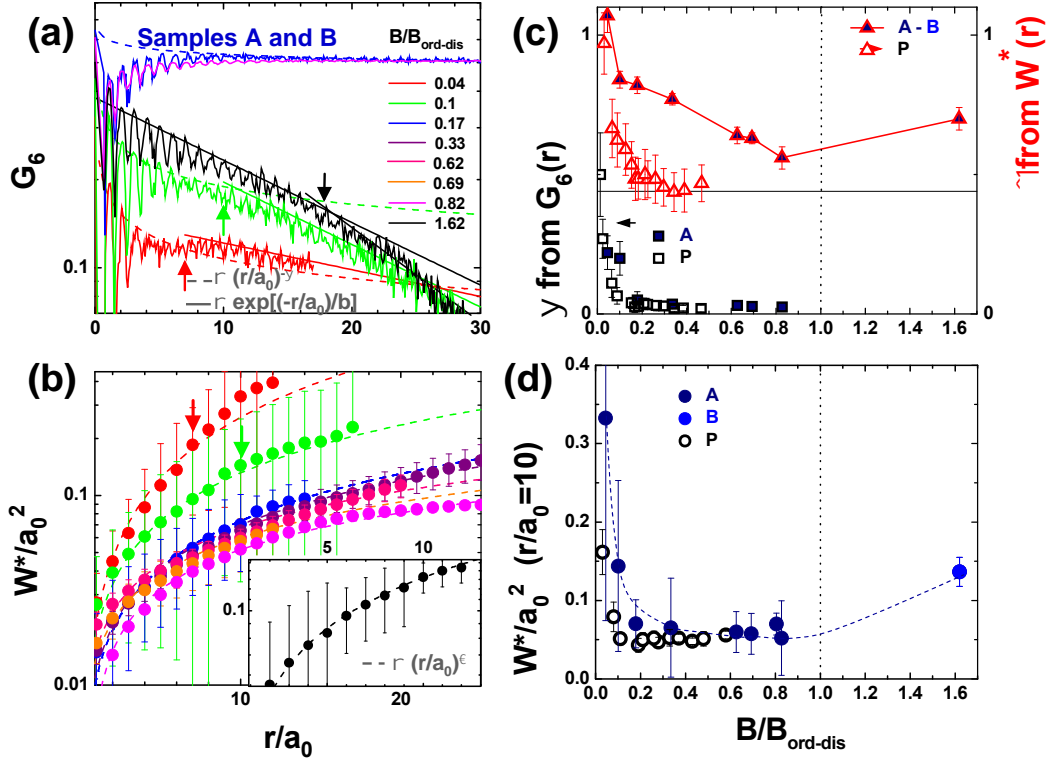}
\caption{Orientational and positional order of the vortex and Bragg
glass phases of Bi$_{2}$Sr$_{2}$CaCu$_{2}$O$_{8 + \delta}$ vortex
matter at the surface of electron-irradiated samples A and B. Data
obtained from magnetic decoration experiments. (a) Orientational
correlation function $G_{\rm 6}$ for various $B/B_{\rm ord-dis}$.
Arrows indicate the size of crystallites. (b) Displacement
correlator $W^{*}/a_{0}^{2}$ calculated using the lanes algorithm
avoiding the effect of topological defects. Data shown following the
same color-code as in (a). Insert: detail of the fit for data
obtained at $B/B_{\rm ord-dis}=1.62$. Dashed lines correspond to
fits with an algebraic decay whereas full lines are exponential
fits. (c) Exponents from the algebraic fittings of $G_{\rm 6}(r)$
(left axis, squares) and of $W^{*}(r)/a_{0}^{2}$ (right axis,
triangles) for the Bragg and vortex glass phases of samples P, A and
B. The dashed red line indicates the characteristic value of
$\nu=0.44$ expected for the random manifold regime. (d)
Field-evolution of the displacement correlator at a fixed lattice
spacing, $W^{*}(r/a_{0}=10)/a_{0}^{2}$. } \label{Figure4}
\end{figure*}

\begin{figure*}[ttt]
\includegraphics[width=0.9\textwidth]{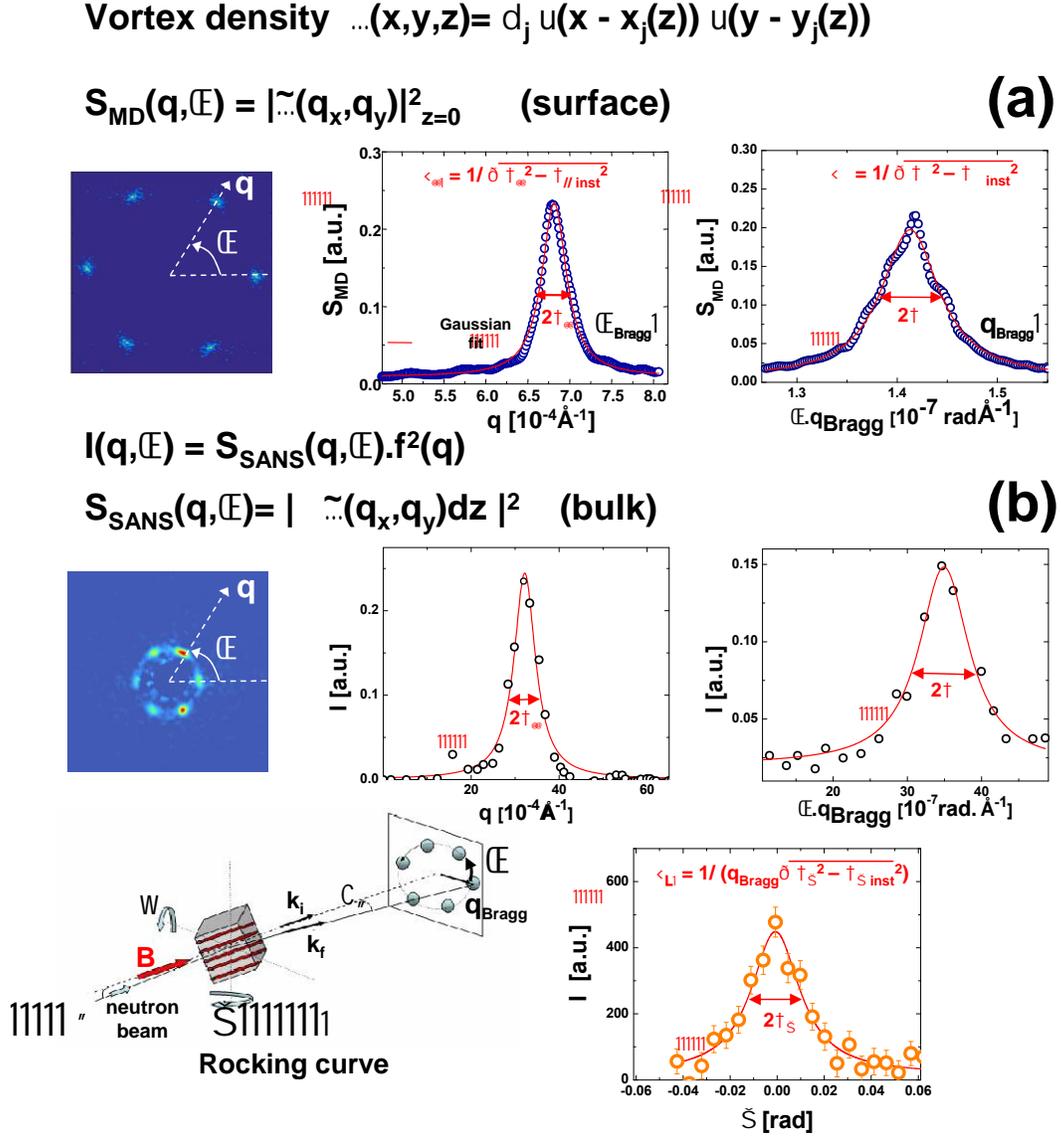}
\vspace{-4.5cm} \caption{In-plane and along vortices correlation
lengths as measured by  small-angle neutron scattering (SANS) and
magnetic decoration (MD) techniques. (a) Left: Structure factor at
the sample surface obtained by imaging vortex positions with MD,
$S_{\rm MD}(q,\Psi)$ (sample P at $B/B_{\rm ord-dis} = 0.15$).
Middle: Radial $q$-profile of $S_{\rm MD}$ at a fixed $\Psi$ on
traversing a Bragg peak and calculation of the in-plane correlation
length along the $q$ direction, $\xi_{\parallel}/a_{0}$. Right:
azimuthal $\Psi$-profile of $S_{\rm MD}$ at $q_{\rm Bragg}$ and
perpendicular in-plane correlation length $\xi_{\perp}/a_{0}$. (b)
Top left: Neutron diffraction pattern at Bragg condition (sample P
at $B/B_{\rm ord-dis} = 1.25$)proportional to the structure and the
magnetic form factors of the vortex lattice $I(q, \Psi)= S_{\rm
SANS}(q, \Psi).f^{2}(q)$ at the whole volume of the sample. Top
middle: radial $q$-profile of $I(q,\Psi)$ on traversing a Bragg
peak. Top right: azimuthal $\Psi$-profile of $I(q,\Psi)$. Similarly
as in the case of MD, fitting each profile with Gaussians yields the
two in-plane correlation lengths $\xi_{\parallel}/a_{0}$ and
$\xi_{\perp}/a_{0}$. Bottom: experimental configuration and typical
rocking-curve data from which the longitudinal correlation length,
$\xi_{\rm L}/a_{0}$, can be exttracted.} \label{Figure5}
\end{figure*}

\begin{figure*}[ttt]
\includegraphics[width=0.9\textwidth]{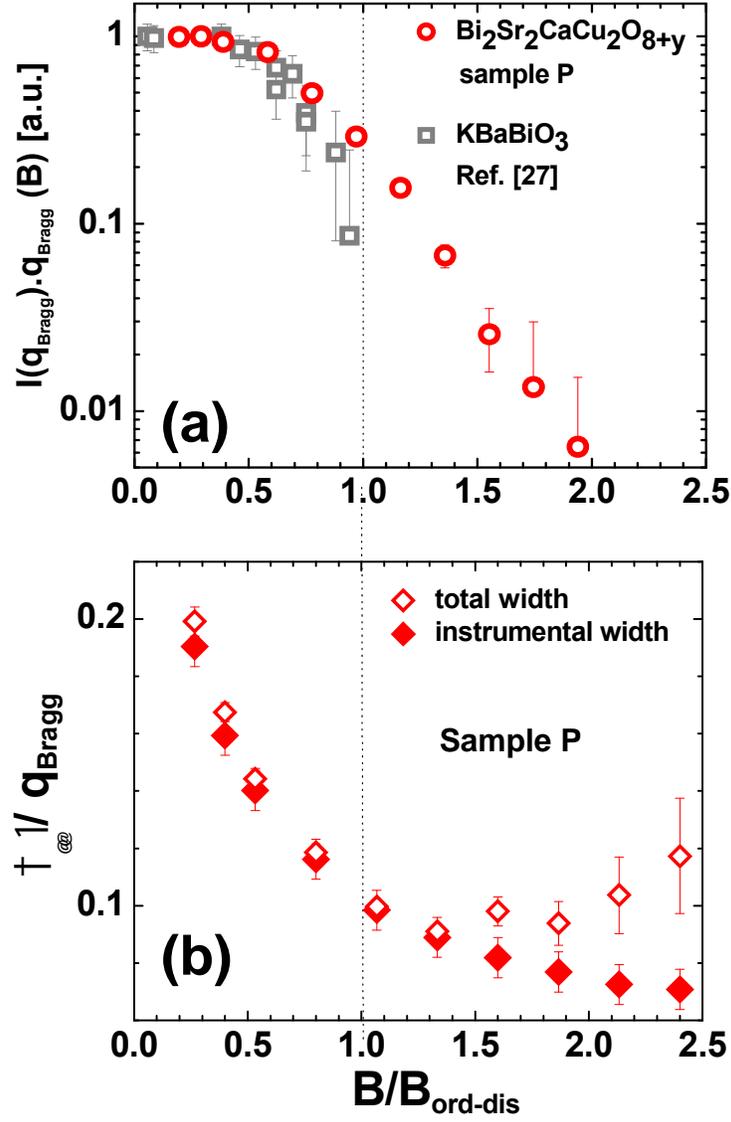}
\caption{Neutron diffracted intensity (at $q=q_{\rm Bragg}$) and
width of the Bragg peaks as a function of $B/B_{\rm ord-dis}$. (a)
Normalized intensity in pristine Bi$_{2}$Sr$_{2}$CaCu$_{2}$O$_{8 +
\delta}$ (circles) compared to similar data in (K,Ba)BiO$_{3}$
(squares) taken from Ref.\,\onlinecite{Klein2001}. (b) Average
$q$-width of the Bragg peaks, $\sigma_{\parallel}/q_{\rm Bragg}$,
for pristine Bi$_{2}$Sr$_{2}$CaCu$_{2}$O$_{8 + \delta}$ in
comparison to our instrumental resolution.} \label{Figure6}
\end{figure*}

\begin{figure*}[ttt]
\includegraphics[width=0.9\textwidth]{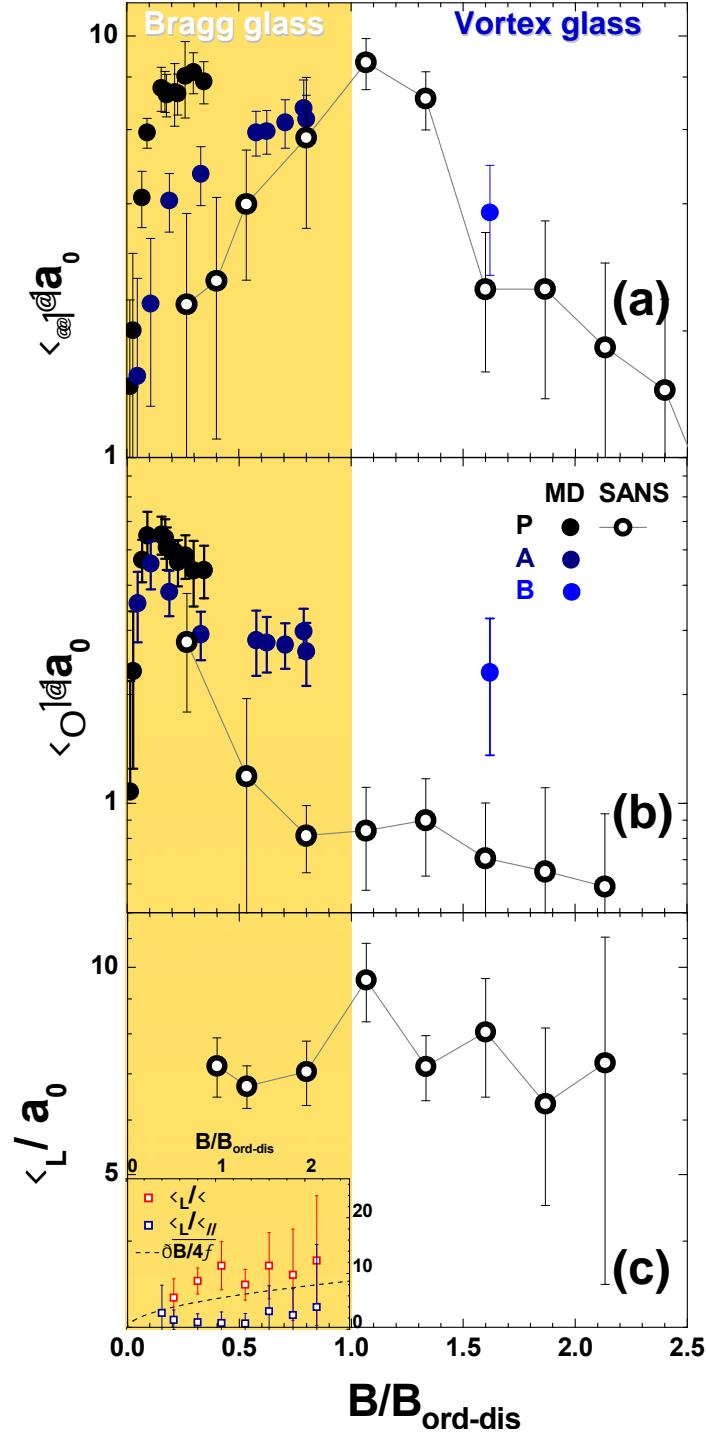}
\caption{ Field-evolution of the in-plane and longitudinal
correlation lengths obtained from magnetic decoration (MD) data
collected at the surface and small-angle-neutron-scattering (SANS)
data probing vortex displacements along the sample thickness. Data
for the Bragg and vortex glass phases in pristine samples P and
electron-irradiated samples A and B. In-plane correlation lengths
(a) parallel to the $q$-direction $\xi_{\parallel}/a_{0}$ and (b)
along the azimuthal direction $\xi_{\perp}/a_{0}$. (c) Longitudinal
correlation length sensitive to the meandering of vortices along the
field direction. Insert: Ratio between the longitudinal and in-plane
correlation lengths obtained from SANS and comparison with the
field-evolution of the bending energy term, $\sqrt{c_{\rm 44}/c_{\rm
66}} \sim \sqrt{B/4\pi}$, expected for
Bi$_{2}$Sr$_{2}$CaCu$_{2}$O$_{8 + \delta}$ (dashed line). }
\label{Figure7}
\end{figure*}

\begin{figure*}[ttt]
\centering
\includegraphics[width=0.9\textwidth]{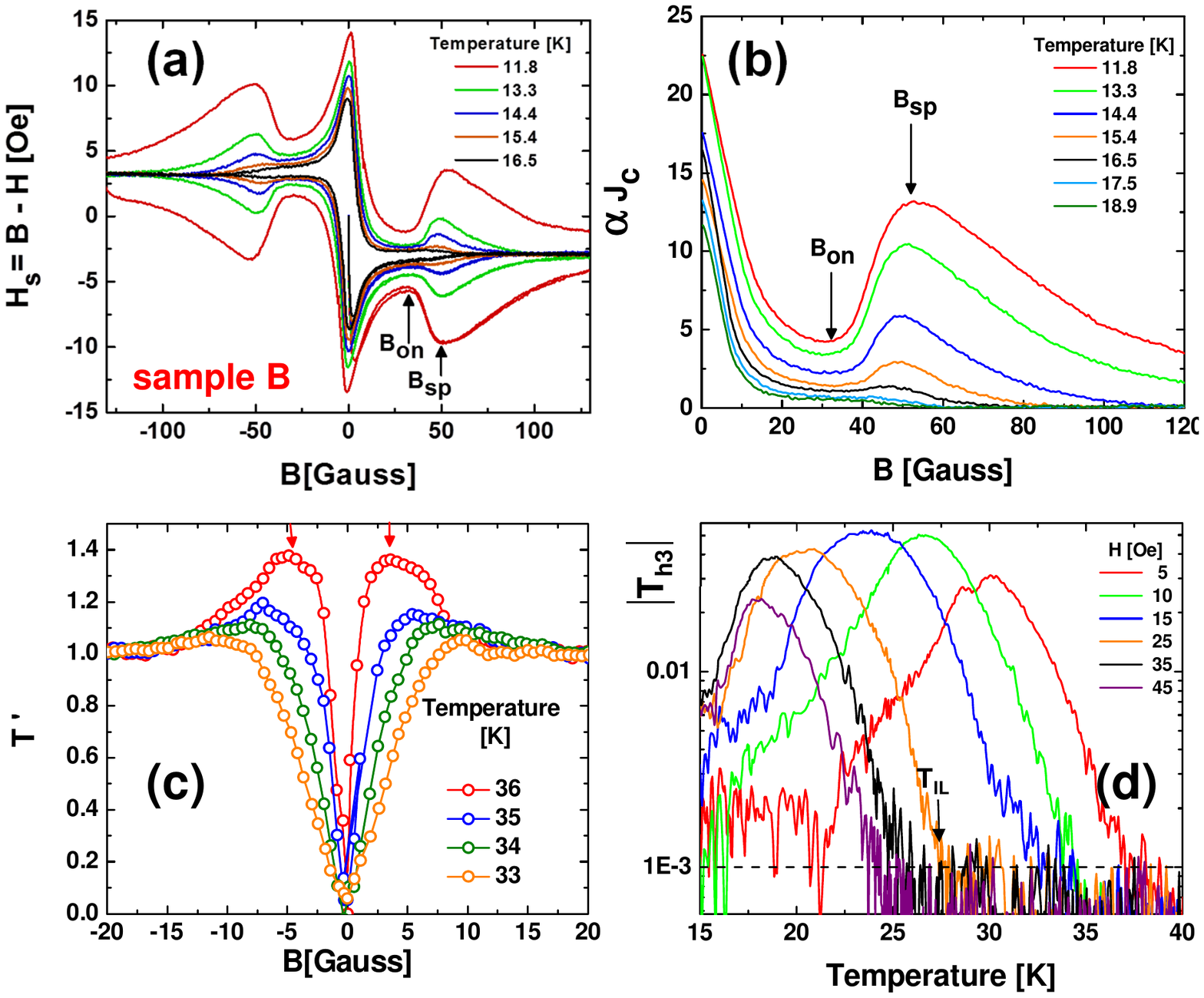}
\vspace{-5 cm} \caption{Detection of the vortex first-order
transitions $B_{\rm FOT}$ (high temperatures) and $B_{\rm ord-dis} =
B_{\rm ON}$ (low temperatures), and the irreversibility line for
Bi$_{2}$Sr$_{2}$CaCu$_{2}$O$_{8 + \delta}$. Data for the
illustrative case of electron-irradiated sample B are obtained from
local dc and ac magnetic measurements. (a) dc magnetization loops at
low temperatures.(b) Separation of the ascending and descending
branches of dc loops, proportional to the critical current $J_{\rm
c}$. Both figures highlight the field for the onset, $B_{\rm ON}$,
and full development, $B_{\rm SP}$, of the peak in critical current,
the former associated with the order-disorder transition. (c)
Transmittivity loops at different temperatures in the
high-temperature region showing paramagnetic peaks  at $B_{\rm FOT}$
(see arrows). (d) Third harmonic signal presenting a non-negligible
value  when the magnetic response becomes non-linear at the
irreversibility line $H_{\rm IL}$ (experimental noise indicated with
a dotted line). ac measurements performed with a ripple field of
1\,Oe in amplitude and frequency of 7.1\,Hz.} \label{FigureSM1}
\end{figure*}

\begin{figure*}[ttt]
\centering
\includegraphics[width=0.9\textwidth]{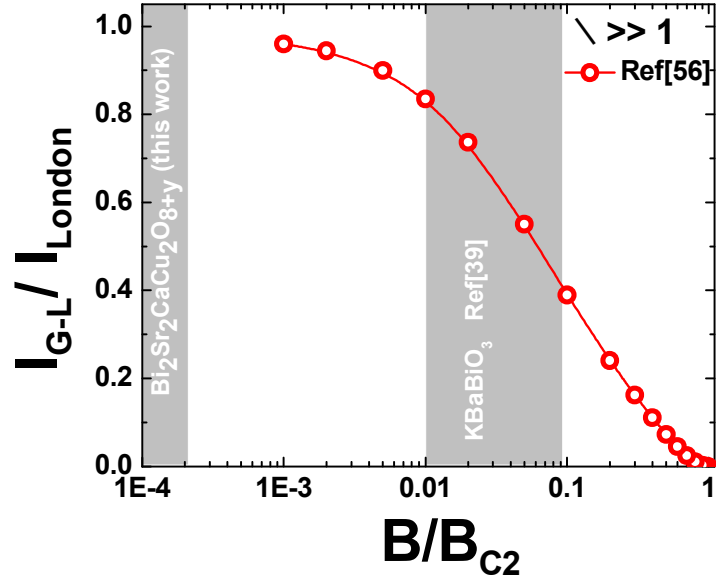}
\caption{Ratio between the intensity of Bragg peaks in the
Ginzburg-Landau and the London approximations, $I_{\rm G-L}/I_{\rm
London}$, as a function of $B/B_{\rm c2}$. This theoretical curve
was calculated from Ref.\,\onlinecite{Brandt1997}. The light-gray
regions indicate the field-ranges studied experimentally in
Bi$_{2}$Sr$_{2}$CaCu$_{2}$O$_{8 + \delta}$ in this work and in
(K,Ba)BiO$_{3}$ in Ref.\onlinecite{Klein2001}.} \label{FigureSM4}
\end{figure*}

\begin{figure*}[ttt]
\centering
\includegraphics[width=0.9\textwidth]{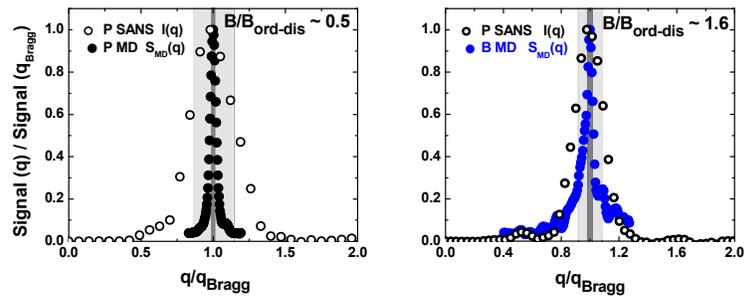}
\caption{Comparison between the $q$-profiles of the normalized
signals as observed by MD and SANS at the same $B/B_{\rm ord-dis}$
below (left 0.4) and above (right 1.62) the order-disorder
transition. The light-gray areas indicate the instrumental
resolution of SANS data, mainly determined by the beam divergence
and wavelength resolution. The dark-gray areas highlight the
instrumental resolution of MD data. } \label{FigureSM5}
\end{figure*}

\begin{figure*}[ttt]
\centering
\includegraphics[width=0.9\textwidth]{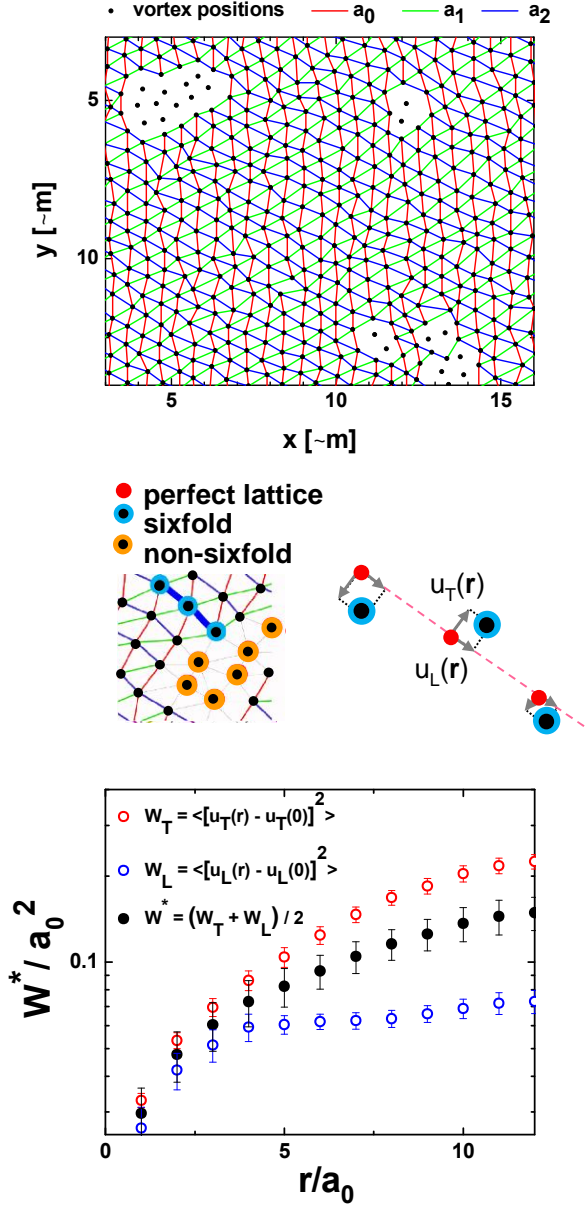}
\caption{Top: Vortex positions and lanes considered to calculate the
local vortex displacements $\mathbf{u(r)}$ with respect to the
lattice sites along every avenue $a_{\rm i}$ (typical structure for
 the Bragg glass phase of the pristine sample P). Middle:
Schematics of the vortex displacements computed to calculate the
displacement correlator $W^{*}(r)$ along ($u_{\rm L}(r)$) and
perpendicular ($u_{\rm T}(r)$) to a given lane at a distance $r$
from its starting point. The positions of vortices are indicated
with black dots with light-blue (orange) borders for sixfold
(non-sixfold) coordinates vortices. The positions corresponding to
the
 triangular lattice are shown in small red dots. Bottom:
Transversal, longitudinal y total displacement correlators for this
vortex structure as a function of $r/a_{0}$.} \label{FigureSM2}
\end{figure*}

\begin{figure*}[ttt]
\centering
\includegraphics[width=0.9\textwidth]{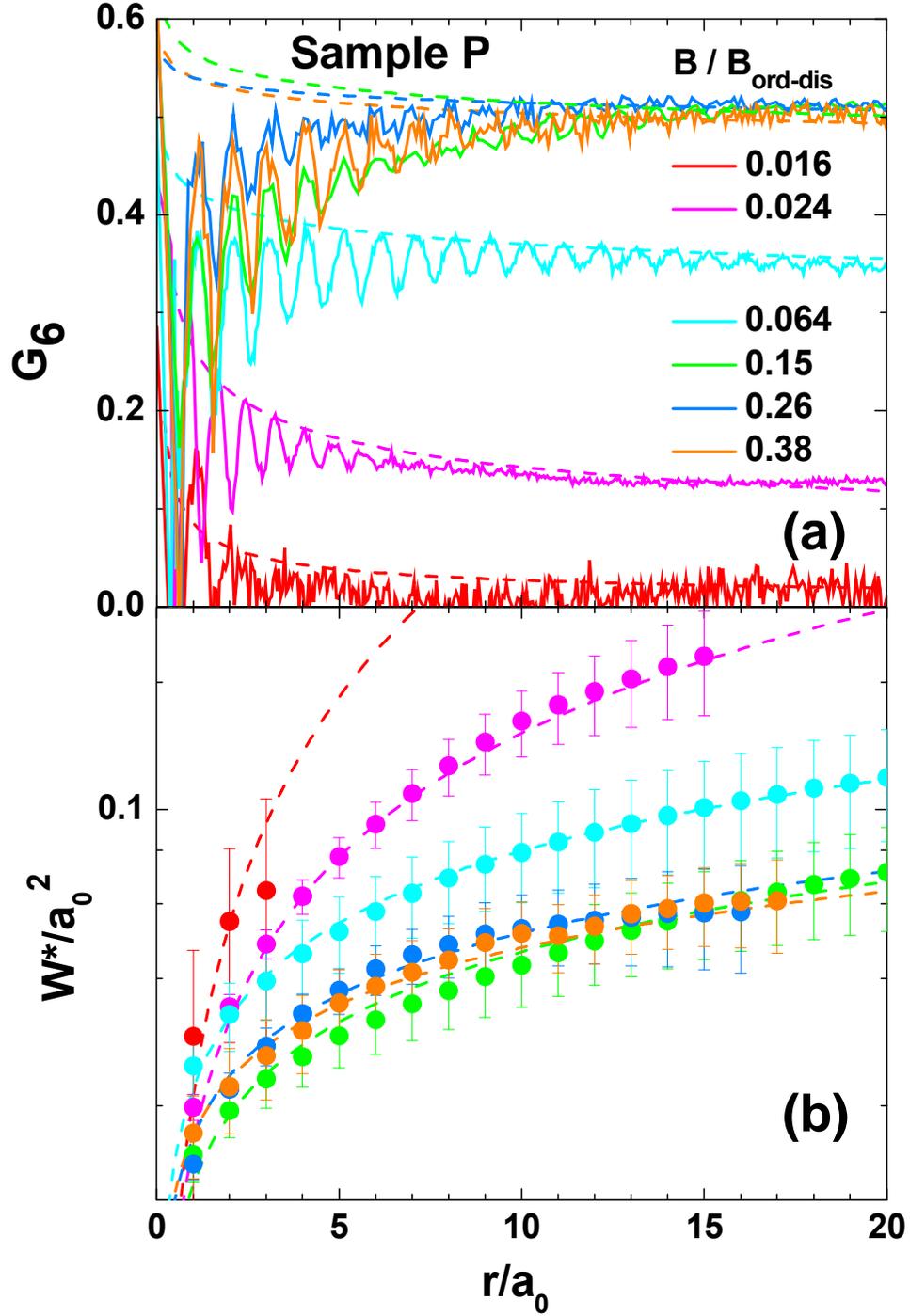}
\caption{Orientational and positional order in the Bragg glass phase
of pristine Bi$_{2}$Sr$_{2}$CaCu$_{2}$O$_{8 + \delta}$ vortex
matter. (a) Orientational correlation function $G_{\rm 6}$ for
various $B/B_{\rm ord-dis}<1$. Dashed lines correspond to fits with
an algebraic decay. (b) Displacement correlator $W^{*}/a_{0}^{2}$
calculated using the lanes algorithm (see text). Dashed lines: fits
with an algebraic growth typical of the random manifold regime of
the Bragg glass phase.} \label{FigureSM3}
\end{figure*}

\end{document}